\pdfoutput=1
\documentclass[journal]{IEEEtrann}
\usepackage{graphicx}

\ifCLASSINFOpdf
\else
\fi
\begin{document}
%
\title{Total and Partial Fragmentation Cross-Section of 500 MeV/nucleon Carbon Ions on Different Target Materials}
%
%
%

\author{Behcet~Alpat,~\IEEEmembership{}
Ercan~Pilicer,~\IEEEmembership{}
Sandor~Blasko,~\IEEEmembership{}
Diego~Caraffini,~\IEEEmembership{}
Francesco~Di Capua,~\IEEEmembership{}
Vasile~Postolache,~\IEEEmembership{}
Giorgio~Saltanocchi,~\IEEEmembership{}
Mauro~Menichelli,~\IEEEmembership{}
Laurent~Desorgher,~\IEEEmembership{}
Marco~Durante,~\IEEEmembership{}
Radek~Pleskac,~\IEEEmembership{}
Chiara~La Tessa~\IEEEmembership{}

\thanks{B. Alpat and M. Menichelli are with Istituto Nazionale di Fisica Nucleare - Sezione di Perugia, Via A. Pascoli, Perugia, 06123, Italy.}
\thanks{E. Pilicer is with Department of Physics, Uludag University, 16059, Bursa, Turkey. During this work he had a research contract with the Istituto Nazionale di Fisica Nucleare - Sezione di Perugia.}
\thanks{S. Blasko, D. Caraffini, F. Di Capua, A. Lucaroni,
  V. Postolache and G. Saltanocchi are with MAPRad S.r.l., Via Cristoforo Colombo 19/I, 06127 Perugia, Italy.}
\thanks{L. Desorgher is with SpaceIT GmbH, 3007  Bern, Switzerland.}
\thanks{M. Durante, R. Pleskac and C. La Tessa are with GSI, Darmstadt, Germany.}

\thanks{This work has been supported under ESA-ESTEC contract number 21985/08/NL/AT.}}

\maketitle

\begin{abstract}
By using an experimental setup based on thin and thick double-sided 
microstrip silicon detectors, it has been possible to identify the
fragmentation products due to the interaction of very high energy
primary ions on different targets.  Here we report total and partial
cross-sections measured at GSI (Gesellschaft f{\"u}r Schwerionenforschung), Darmstadt, for 500 MeV/n energy
$^{12}C$ beam incident on water (in flasks), polyethylene, lucite,
silicon carbide, graphite, aluminium, copper, iron, tin, tantalum and
lead targets. The results are compared to the predictions of GEANT4 (v4.9.4)
and FLUKA (v11.2) Monte Carlo simulation programs.
\end{abstract}

\begin{IEEEkeywords}
Heavy ions, shielding materials, GEANT4, FLUKA, charge-changing, nuclear fragmentation  cross-section. 
\end{IEEEkeywords}

%
\IEEEpeerreviewmaketitle

\section{Introduction}
%
%
%
%

The nuclear interaction mechanism of carbon ions with nuclei represent
a key point in the understanding of delivered dose in cancer therapy applications. The effect of fragmentation diminishes the number of
primary ions delivered to the region under treatment and produces a
tail of damaging ionisation beyond the Bragg peak.  In addition, the
trajectories of fragments may be sufficiently perturbed compared to
that of the incident ion that they deposit a non-negligible dose
outside the volume of tissue being treated.  The nuclear fragmentation studies are of
 interest also for radiation effects community \cite{clemens,sabra}  since both primary Galactic Cosmic Rays (GCR) or 
their secondaries may cause Single Event Effects (SEE) on electronics systems and instrumentations. Carbon ions are  a
significant component of the  GCR and of Solar Particle Events (SPEs) (see for instance  \cite{mxapsos} for a 
review of models of the near-Earth space radiation environment) and 
they contribute to the dose delivered to astronauts as well as to SEE occurring in electronics on long duration
space missions. The external environment is fairly well known, but
when the incident radiation interacts with the spacecraft hull and
internal materials, nuclear fragmentation modifies the external
radiation field \cite{wilson}.  A recent exhaustive review on this
subject are given in \cite{durante,silvestri1,silvestri2}.  Both applications, cancer therapy
dose and optimization of shielding materials for space explorations,
require an accurate knowledge of the physics of the fragmentation
process. In this work we measured the total charge-changing
cross-sections and partial cross-sections for individual fragments
produced by a 500 MeV/n carbon beam on eleven different target
materials. The experimental results are compared to the predictions of
GEANT4 \cite{geant1,geant2} and FLUKA \cite{FLUKA1,FLUKA2} Monte Carlo
simulation programs.

\section{Experimental Setup}
\begin{figure}
\centering
\includegraphics[width=3.5in]{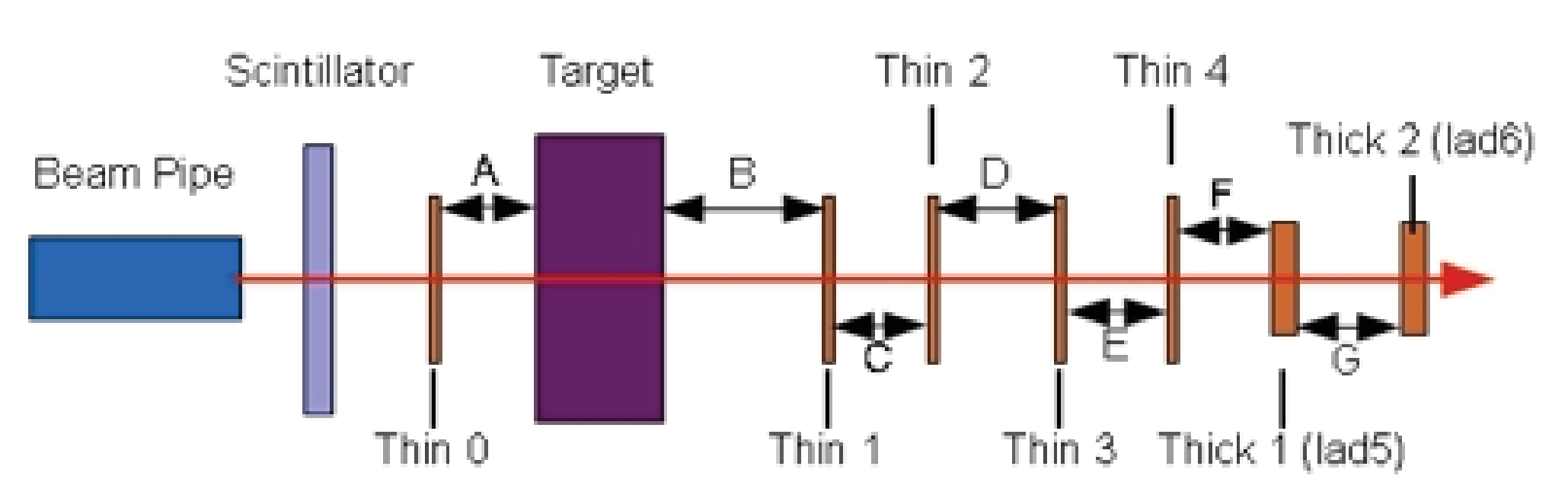}
\caption{\label{fig:sketch} A general sketch of the setup: the
  position of Thin and Thick detectors}
\end{figure}

The detection system designed and assembled by
MAPRad \cite{maprad} at INFN (Istituto Nazionale di Fisica Nucleare) Perugia clean room
facilities \cite{clroom}.
The experimental apparatus (see Fig.~\ref{fig:sketch}) used to
perform the measurement is based on the double-sided microstrip
silicon detector (DSSD) technology.

Our first test set consisted of five double-sided $300~\mu$m thin microstrip
silicon detectors (DSSD300) of the same type used in the AMS02
experiment \cite{ams2}.

The implants on the two sides of DSSD300 run along orthogonal
directions, providing X and Y measurements of ion impact position, with a
readout pitch of 0.44 mm and external dimensions of
41x72x0.31 mm$^2$.  The number of channels connectable to Front-End
Electronics (FEE) is 160 + 96 for p and n side respectively however, on both sides, only
central 96 channels were connected to FEE. 

The second test set included two 1.5 mm thick microstrip silicon
detectors (DSSD1500), produced by FBK of Trento (Italy), with
64 + 64 channels on p and n sides respectively. These sensors have a
readout pitch of 0.5 mm and external dimensions of 35x35x1.5 mm$^2$
and all channels on both sides were connected to their respective FEE.  The
detectors mentioned above were arranged with the first (Thin0) and the
second (Thin1) of the thin sensors, placed in fixed positions before
and after the target, and were followed by the other three sensors
(Thin2, Thin3, Thin4).  The two thick sensors (Thick1, Thick2) were
positioned as the last two sensors of the experimental
setup. Fig.~\ref{fig:sketch} shows a schematic drawing of the setup
on the beam line for the ions.

The positions of Thin0 and Thin1 were chosen so that any target
thickness among those available could be accomodated in between.  The
positions of the other sensors were varied during the experiment to
assess the effect of geometrical acceptance on the final result.  The
value of the distances, measured with the laser device with
resolution better than 0.1 mm, are show in Table I. 

\begin{table}[!ht]
\label{tab:dist}
\begin{center}
\caption{Distances (mm) between detectors in the different
  configurations in August 2010 session. For distances corresponding
  to each letter please refer to Fig.~\ref{fig:sketch}}
\begin{tabular}{|l|l|c|c|c|c|r|}
\hline
\hline
 Config. & A+B+Target   &  C  &  D  &  E & F & G\\
\hline
\hline
 1  & 719.33   &  67.96  &  65.56  &  75.28 & 74.80 & 79.61\\
\hline
 2 & 719.33   &  65.59  &  64.23  &  73.17 & 70.80 & 76.79\\
\hline
 3 & 719.33   &  79.39  &  119.66  &  79.45 & 119.29 & 80.25\\
\hline
 4 & 719.33   &  79.56  &  120.76  &  80.67 & 118.69 & 79.49\\
\hline
\hline
\end{tabular}
\end{center}
\end{table}

\begin{figure}
\centering
\includegraphics[width=3.5in]{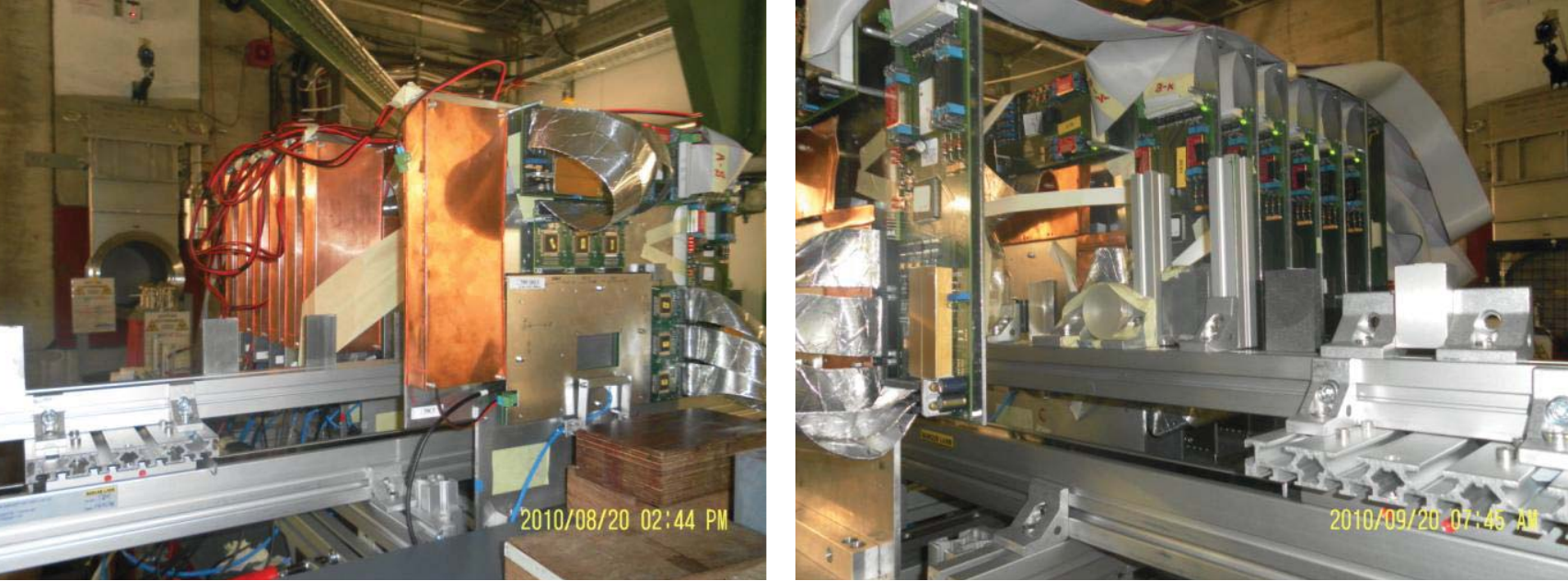}
\caption{\label{fig:setup} The photos show the detector setup at GSI}
\end{figure}

Although the sensors have self triggering capabilities, a plastic
scintillator was inserted between the point where the beam is
extracted in air and the experimental setup, to act as an external
trigger for the data acquisition (DAQ)  electronics.  The setup was equipped with
temperature sensors and with a cooling system, flowing air at
$2\,^{\circ}{\rm C}$ on the surface of the detectors.
Fig.~\ref{fig:setup} shows the experimental setup assembled at GSI.

Testing took place in experimental Cave A at GSI-Darmstadt in August
2010. The beam was $^{12}C$ with an energy of 500 MeV/n. Beam
intensity was about 5k to 10k counts per spill with a duration of 10s, according to GSI monitoring system.  Several target materials were
used: water (in flasks), polyethylene, aluminium, lucite,
graphite, iron, tin, lead, copper, tantalum, silicon carbide. For some
target types, several different thicknesses were used. The details are
reported in Table II. 

\begin{table}[!ht]
\label{tab:thickness}
\begin{center}
\caption{Tested target materials and relative thicknesses.}
\begin{tabular}{|l|c|}
\hline
\hline
 Material & Thickness $(g/cm^2)$\\
\hline
\hline
air & 0.035\\
\hline
water & 5.700\\
\hline
lucite & 3.840\\
\hline
polyethylene & 0.465\\
\hline
silicon carb. & 1.580\\
\hline
aluminium & 5.450\\
\hline
graphite & 1.734\\
\hline
graphite & 3.434\\
\hline
copper & 4.628\\
\hline
iron & 3.935\\
\hline
iron & 6.611\\
\hline
tin & 3.655\\
\hline
tin & 5.848\\
\hline
tantalum & 3.330\\
\hline
tantalum & 6.660\\
\hline
lead & 3.689\\
\hline
lead & 6.810\\
\hline
\hline
\end{tabular}
\end{center}
\end{table}

For each target we collected at least $10^6$ events. A set of
data with no target was also taken to account for the contribution due
to interaction in air or in the silicon sensor itself. Data with no target were also used to
evaluate the corrections to be applied to data described in following
sections.

\section{Data Processing}

The data analysis consists of two main phases. During the first phase
a preselection of events that were suitable for the analysis and
evaluation of correction factors on raw charge signals collected in
the sensors were carried out. The second phase includes the track
fitting and particle identification procedures.

For each run, a set of data is collected by using random trigger (with
no beam on) to calculate the pedestal and random noises of readout
channels as well as the common mode noise which is the mean of the
uniform shift over all channels of given readout chip. After random
trigger data the beam was turned-on to collect the real beam data. The
cluster formation, which is the group of adjacent strips collecting
charge created by a traversing particle that survives several
selection cuts were applied.

\subsection{Data Unfolding}

The requests in this stage were mostly geared towards a general quality
of the event and observed signals (clusters of neighbouring readout
channels). For this purpose we considered only those events that are
separated in arrival time with more than $100 \mu$s from the
previous.

The collected charge (Analog to Digital Converter (ADC) counts) on the several sensors has a
noticeable dependence on the particle impact position and on the
different sensors yield. Therefore corrections to these effects are
needed for accurate identification of the fragmentation products.

\subsubsection{Gain Correction}

Both sides of detectors were read-out through a VLSI (Very Large Scale Integration) \cite{ideas} that
consist into self-triggering, pre-amplifier, sample \& hold,
multiplexer circuitry with a dynamic range of $\pm$132 MIPs. The
circuitry has internal calibration line for each of its 32
channels. All VA32TA3.2 were calibrated between few MIPS up to well
above saturation levels by injecting, on bench, known values of charge
load to trace for each channel the calibration curves. The gain
responses of each channel of a given VA32TA3.2 were normalized to its
average value. The same way, average gain of a given side of a given
sensor was then calculated and single channel correction factors were
finally determined.

\subsubsection{$\eta$ Correction}

The parameter $\eta$ is defined using the two channels of a cluster
that collected the highest signals as:
\begin{equation}
\eta=\frac{Q_r}{Q_l+Q_r}
\end{equation}
where the subscripts Ã¢â‚¬Å“lÃ¢â‚¬Â and Ã¢â‚¬Å“rÃ¢â‚¬Â label the collected charge (Q) of the
leftmost and rightmost of the two most highest strips respectively. An
ion impinging close to a readout strip will yield a value of $\eta $
close to 0 or to 1, while ions hitting in the interstrip region will have
$\eta$ close to 0.5.  Due to the poor capacitive coupling towards the
connected readout channels, the hits in the 0.5 region register a
lower charge: while this is not a problem on Thin0, where we have only
primary ions, it tends to mix the different populations of fragments
in the subsequent sensors, so it is necessary to implement a
correction factor depending on $\eta$ to normalize the collected
charge to a common baseline. The correction factor for collected charge in a given $\eta $ bin is the same for all (primary and fragments), hence, the correction was evaluated using data
from a run without target, then it as applied to all other samples.

\subsubsection{Sensor to Sensor Correction}

A further correction was applied to the collected charge to account
for the different yield of each sensor of the setup. Besides the
obvious difference between thin and thick detectors, that implies
different amplitude of the signals, each sensor DAQ combination has a
different yield. The value of charge for primary particle peak from
the various sensors can vary substantially, and this is also true for
the charge produced by fragmentation products; in order to allow for a
uniform treatment of data from different silicon sensors we fitted the
main peaks of each sensor with a gaussian and used the ratio between
the fit's most probable values to scale the charge of thin sensors to
a common base, which we chose to be the value of charge observed on
Thin0. The correction factor was evaluated on a dataset without target
and then applied to all other data sets.

\subsection{Track Fitting}
Each sensor gives the X-Y impact position of the particles, while the
Z coordinate is known a priori from the mounting position of the
sensor.

The position identification capability of DSSD is used in the
subsequent analysis, since it is to be expected that the ions that
pass through the target without fragmentation, will produce a sequence
of hits in the various sensors that lie on a path that is closer to a
straight line.  Therefore fitting the tracks to a straight line and
imposing a cut on the quality of the fit helps to identify both the
unchanged beam ions and the fragments. In addition, track fitting
helps in rejecting events with low spatial correlation between
signals, due to inefficiencies of the detectors, fluctuations in the
measurements and signal pileup when a primary particle
enters the set up during elaboration of signals.

To ensure that we can apply the track fitting, we checked that each
sensor contains at least one signal on the p side; in case more than
one cluster was present, we only consider the one with the highest
charge. Since it is possible that a detector misses the particle for
some inefficiency of the acquisition chain, the number of track points
used in the fits can vary; in particular, it is possible for the track
not to reach the last detector (Thick2) or, while still reaching it,
to have missing intermediate points (holes). For our analysis we
decided to consider only tracks with no holes, but since in principle
a second interaction in the silicon sensors is possible, we allowed
the track to be as short as two track points.

\begin{figure}
\centering
\includegraphics[width=1.75in]{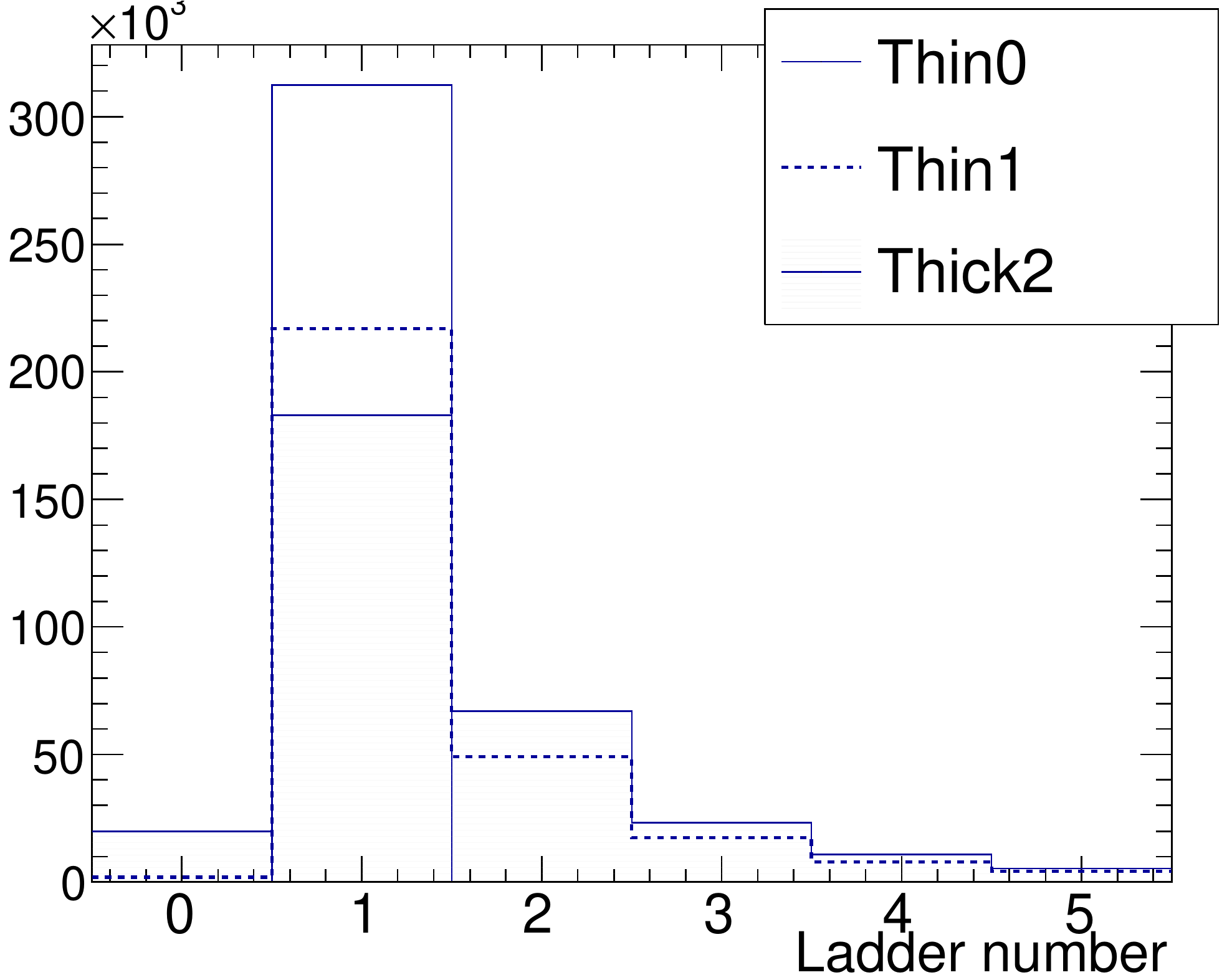}%
\includegraphics[width=1.75in]{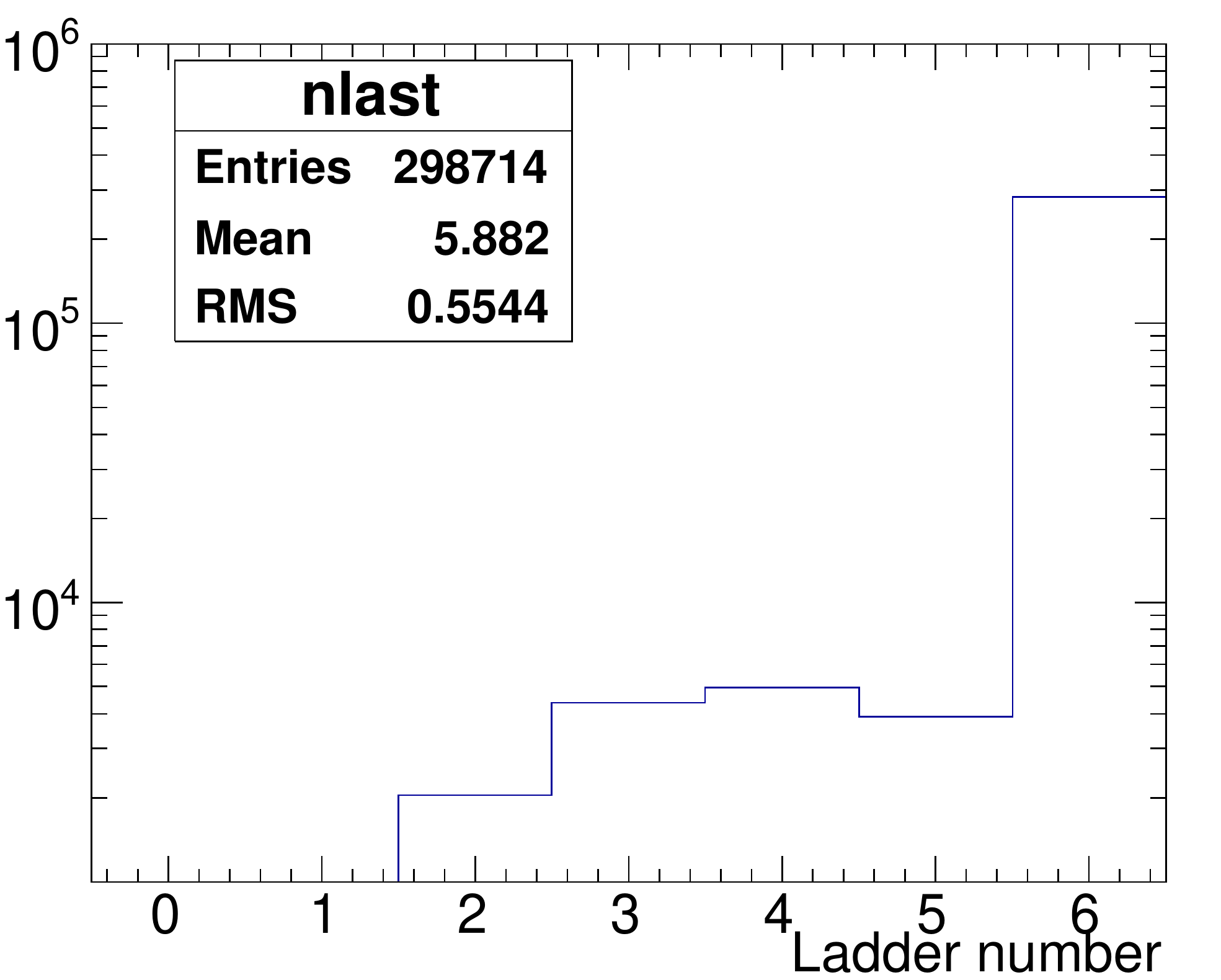}
\caption{\label{fig:seg} Left: distribution of the last detector with
  signal in a given track with no holes. Right: distribution of number
  of cluster per event  per sensors Thin0, Thin1 and Thick2}
\end{figure}

\begin{figure}
\centering
\includegraphics[trim=0cm 0cm 5cm 17cm, clip=true, width=3.5in]{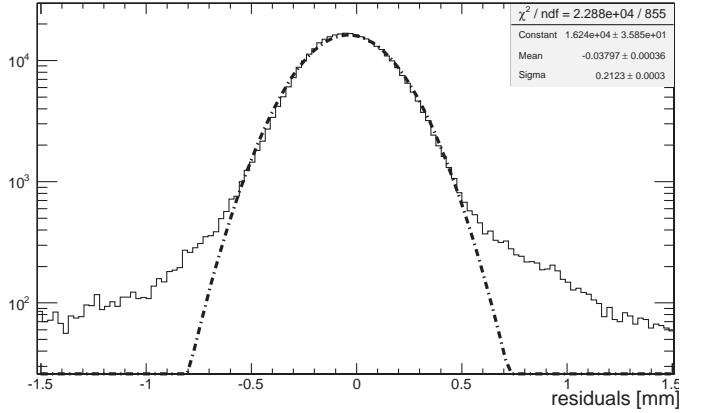}
\caption{\label{fig:resol} Residuals from the actual hit positions and
  fitted tracks for p-side of DSSD300 sensors after the target.}
\end{figure}

Fig. ~\ref{fig:seg} shows, on the right, that the distributions of
the number of clusters for event is peaked at 1, even for the 
most distant sensor (Thick2). On the left side of the same Figure is shown
the distribution of the last sensor with signal without having holes
along the track.  The majority of events reaches the last sensor,
which, considering the small external dimensions of thick sensors, is
consistent with the assumption that the fragments are mostly directed
at low angles from the impinging ions momentum.
Fig.~\ref{fig:resol} shows residuals between fitted track and actual
hit positions on p-side for DSSD300 after the target.

It is worth to notice that the mean value for the absolute residual of
the fits after target is about $210 \mu$m, that is very close to the
value expected for a readout pitch like the one we are using.

All the necessary cluster information, as well as the quantities
calculated in the track fitting procedure are stored in a file, that
is the basis for the following step.

\subsection{Event Selection and Particle  Identification}

To distinguish between the primary ions and the products of the
fragmentation, besides the track fitting results, we use charge
information.

\begin{figure}
\centering
\includegraphics[width=3.5in]{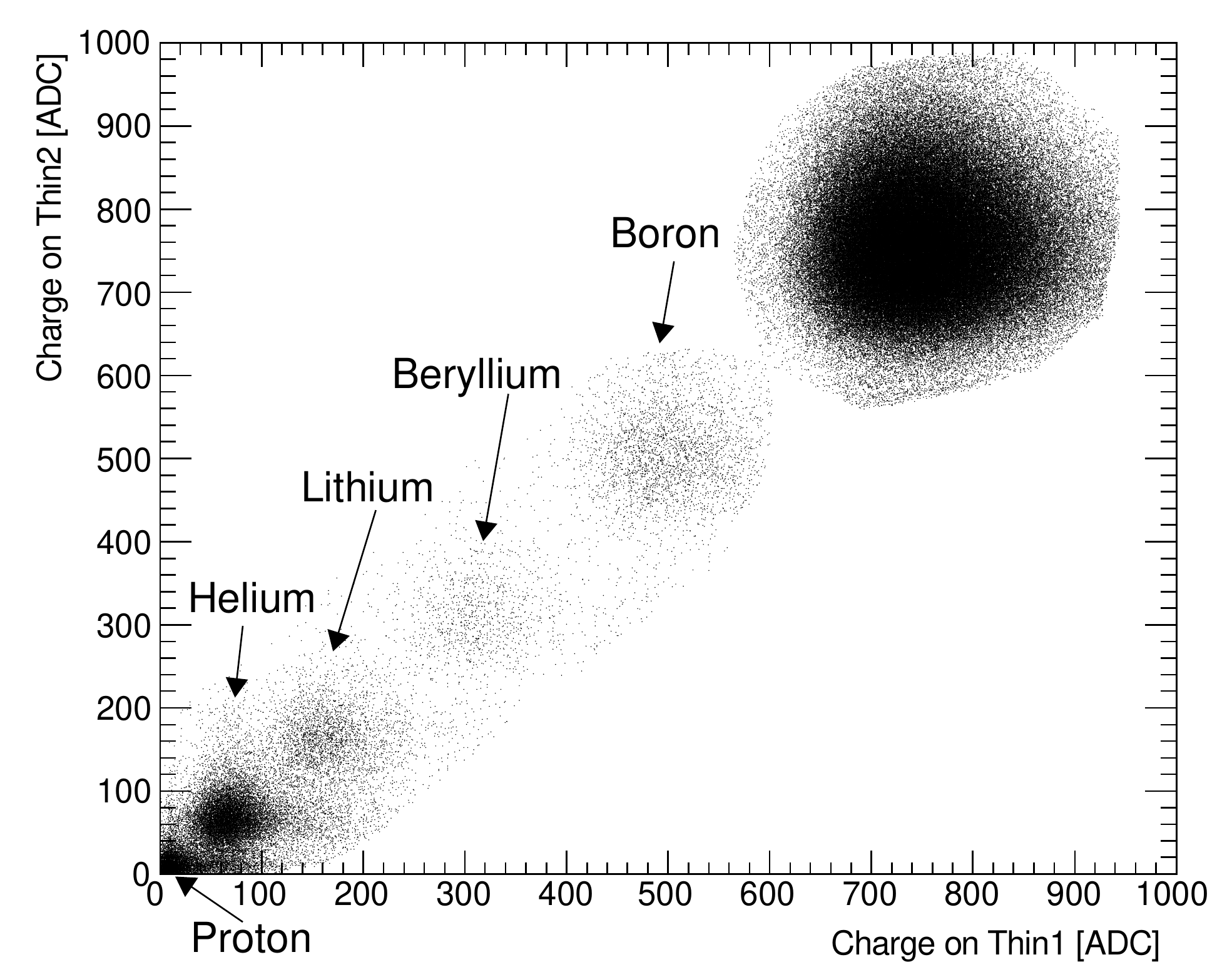}
\caption{\label{fig:charge} Scatter plot of charge observed in
  detector Thin2 versus detector Thin1. Along the diagonal are visible the
  populations due to primary ions and the fragmentation products.}
\end{figure}

 Fig.~\ref{fig:charge}
shows the cluster charge correlation plots of the signals recorded on a pair of thin
sensors (p sides). The plot refers to a pair of sensors (Thin1 and
Thin2), but the same type of plot was produced for all other couples
of sensors of the same kind.

\begin{figure}
\centering
\includegraphics[width=1.75in]{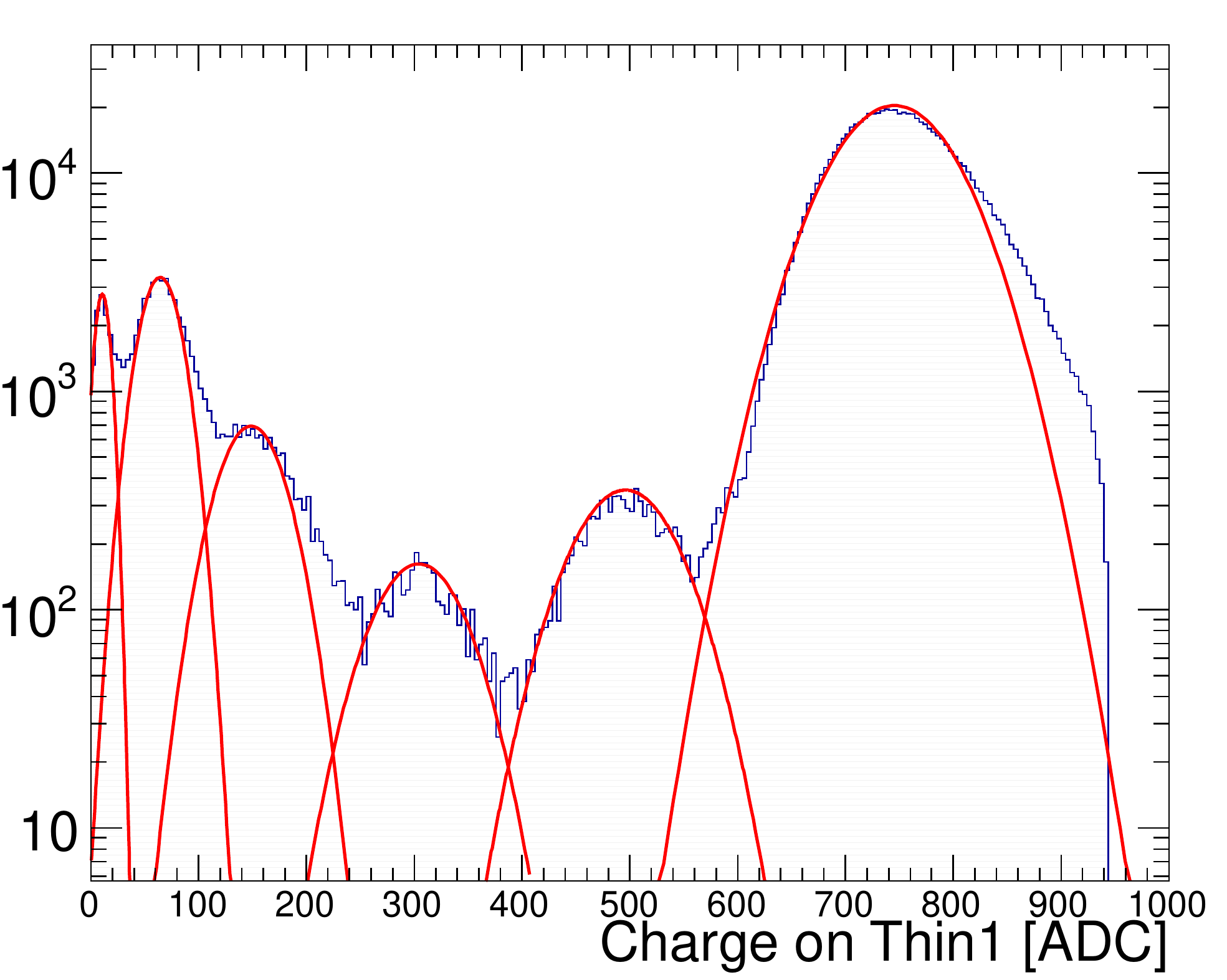}%
\includegraphics[width=1.75in]{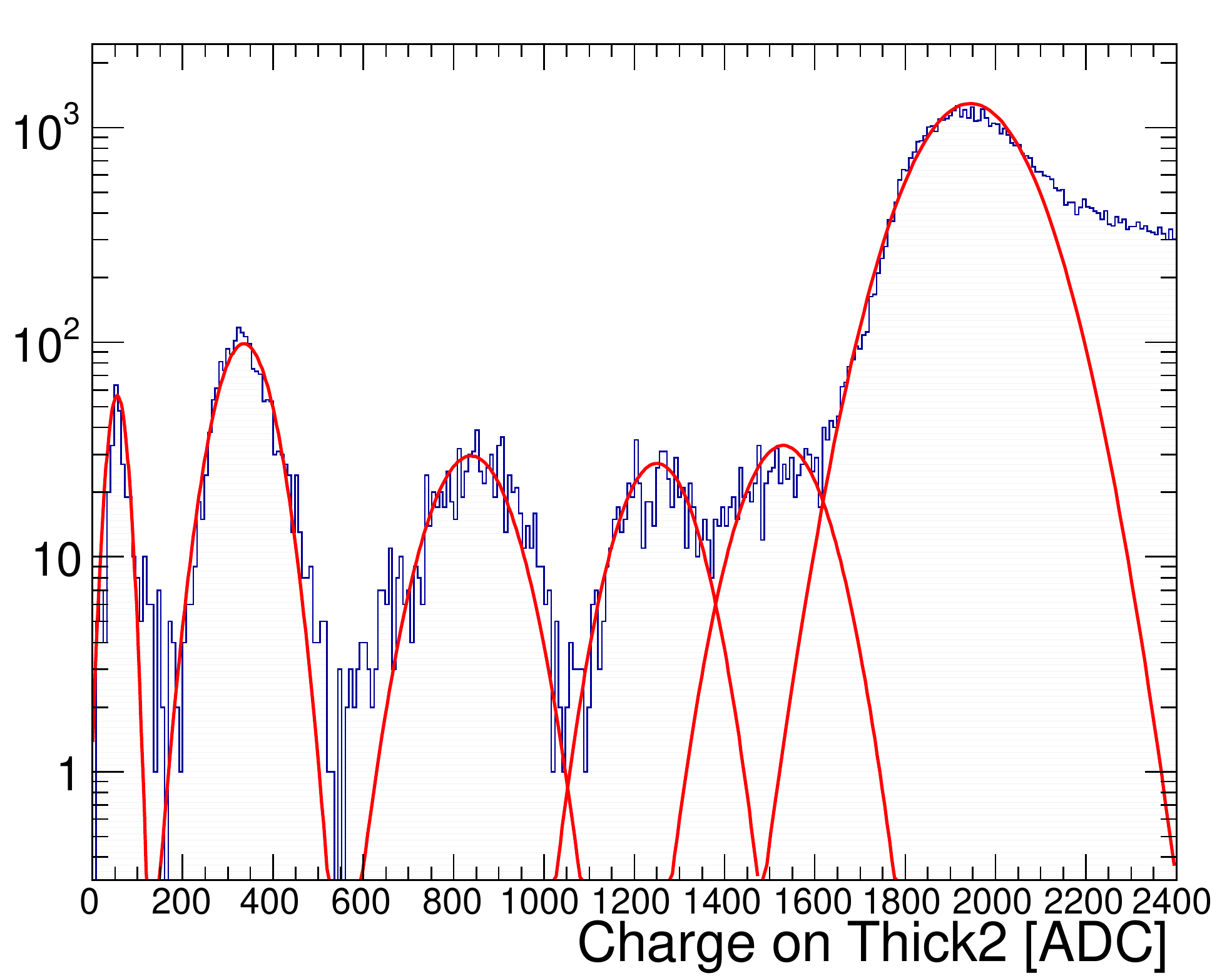}
\caption{\label{fig:projcharge} Collected charge spectra. The left
  plot refers to the first thins after the target, the right one to
  the last thick sensors. The relative positions of the peaks are
  compatible with the atomic number of the fragmentation products and
  the primary carbon beam.}
\end{figure}

 We saw that every couple of sensors basically shows the same
 structure, with a series of approximately round areas aligned along
 the diagonal, which can easily be interpreted as the fragmentation
 products plus the primary peak.  Events with charge signal out of
 diagonal may arise from a series of different reasons, such as
 inefficiencies in the detector's charge collection. Such events was
 removed with a graphical cut. The application of the graphical cut
 yields a cropped scatter plot for each of the sensor combination
 mentioned above. From each one of these cropped plots we can obtain
 two collected charge spectra by projecting the points to one of the
 axes (Fig.~\ref{fig:projcharge}).  The vertical bands are
 representing one sigma limits around most probable charge values. For
 their use see LR(Z) calculation in next paragraph.

  In this analysis, in order to reduce the charge misidentification due
  to overlap in charge distributions, in spite of loss of statistics,
  we considered only the events wich falls within one sigma of the
  charge distribution peaks.
 
\begin{figure}
\centering
\includegraphics[width=3.5in]{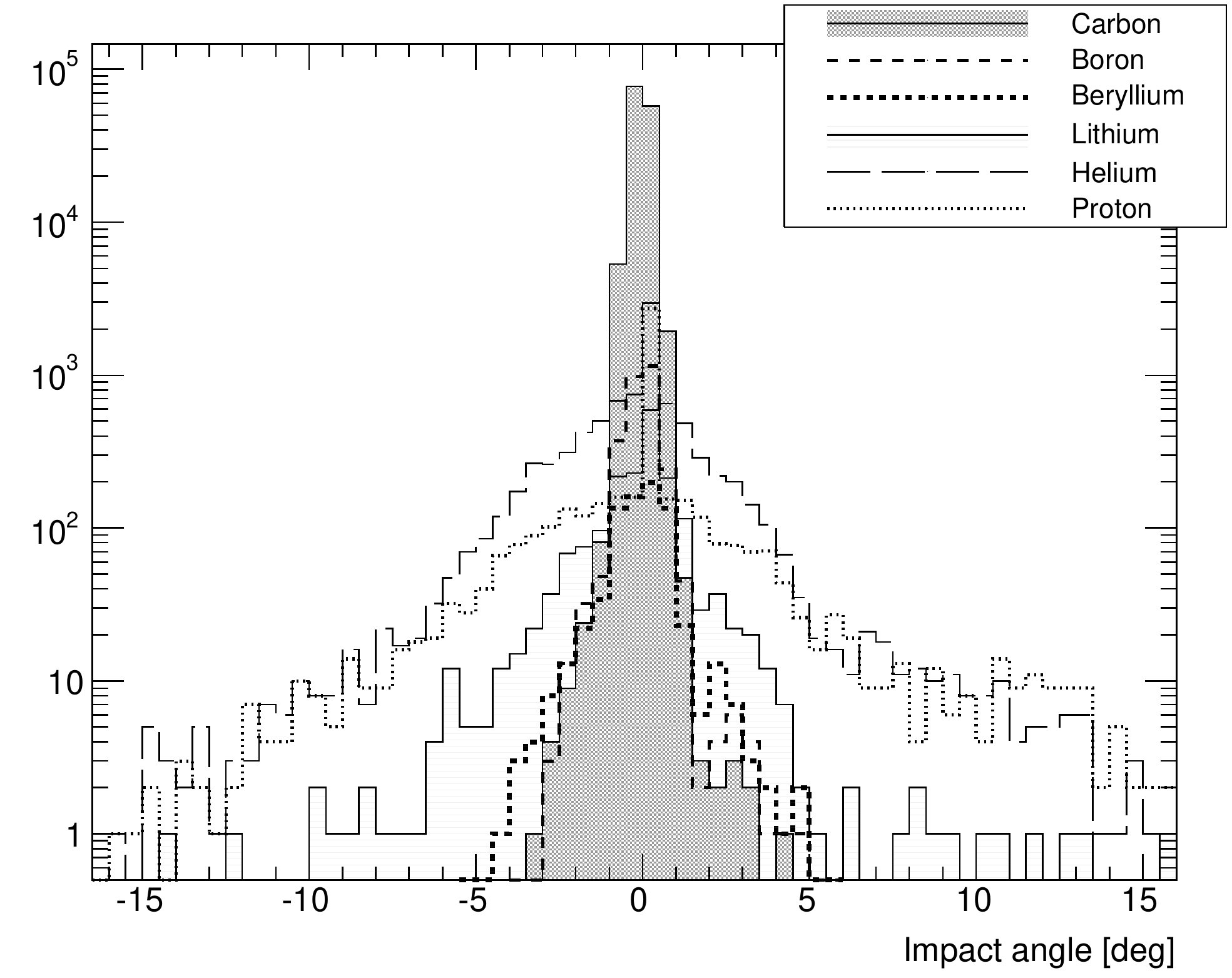}
\caption{\label{fig:slope} Angles as evaluated by track fitting
  procedure for different charge populations. Low Z ions scatter off
  the target at higher angles.}
\end{figure}

Fragments of different charge have different track fit quality, in
particular, for low charge fragments, up to Li, fit quality is worse
possibly because in their case, fragment multiplicity increases,
leading to a mixing of tracks and degraded fitting performance. By
selecting the events that are close to the peaks in the various charge
spectra, we observed that the higher the charge released in the
sensors, the smaller the deviation from the primary ion trajectory and
the better is the fit quality (Fig.~\ref{fig:slope}).

A charge dependent cut on the Chi-square fit probability has been
applied: we required a fit probability greater than 0.9 for events
with ADC counts more than 500, a fit probability greater than 0.8 for
events with falling in the interval between 250 and 500 ADC counts and
no cut on fit probability for events with less than 250 ADC counts.

This analysis procedure allows to obtain a set of fitted peak functions
for each sensor. Such sets of function can be used to implement a
Likelihood test. To this purpose we used the fitted parametrisation
$f_{Z_{S}}(x)$ as a function of the value of charge for a given
fragment species of charge Z and measured on a given sensor S, to
define a normalised probability density function:

\begin{equation}
P_{Z_{S}}(x)=\frac{f_{Z_{S}}(x)}{\int f_{Z_{S}}(x)dx}  
\end{equation}

For each event, we can combine the probability that signal of amplitude $x_{S}$ from sensor S to be the effect of ion Z by multiplying them. We define the LogLikelihood that an event is due to ion of atomic number Z as:

\begin{equation}
L(Z)=-\sum{log(f_{Z_{S}}(x_s))}
\end{equation}

We can then compute at each event the Likelihood for the various Z values and choose the most suitable. To perform the identification we define a test quantity:

\begin{equation}
LR(Z)=\frac{L(Z)}{L(Z)+L(Z-1)}
\end{equation}

Here LR(Z) stands for Likelihood Ratio: this test is specifically
targeted at the discrimination of one peak from the ones with Z
differing by one unit, since the major contribution to wrong charge
assignments are essentially coming from neighbouring peaks.  To study
the capability of the test to identify the ions we performed a very
strict selection, containing only the events whose collected charge
was within one sigma of the same peak in Thin0 and Thin4 and
considered them as “pure samples”. In this way we can evaluate the
LR(Z) with a fairly accurate a priori knowledge of the real ion specie
and see if it is identified as such.

\begin{figure}
\centering
\includegraphics[width=1.8in]{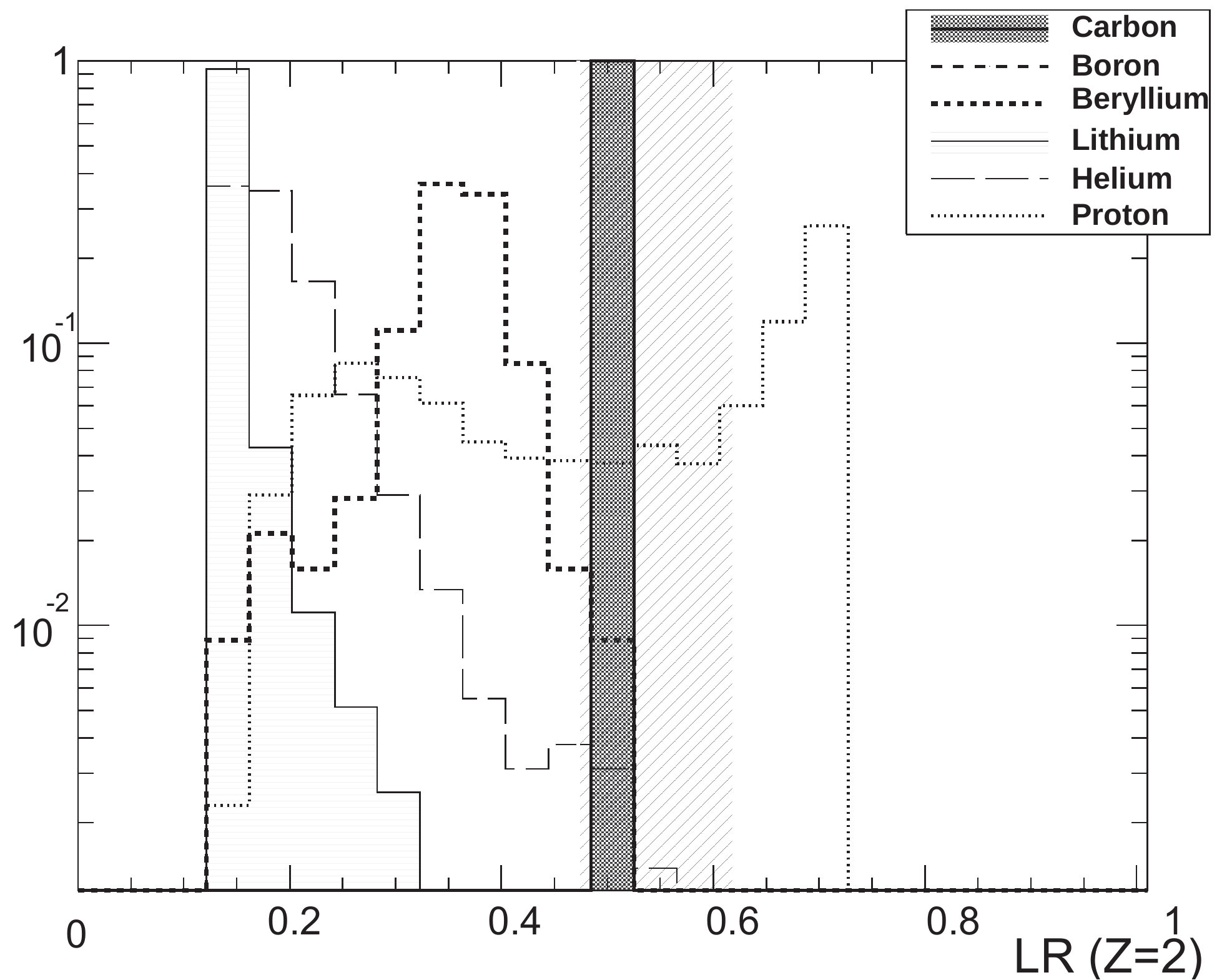}%
\includegraphics[width=1.8in]{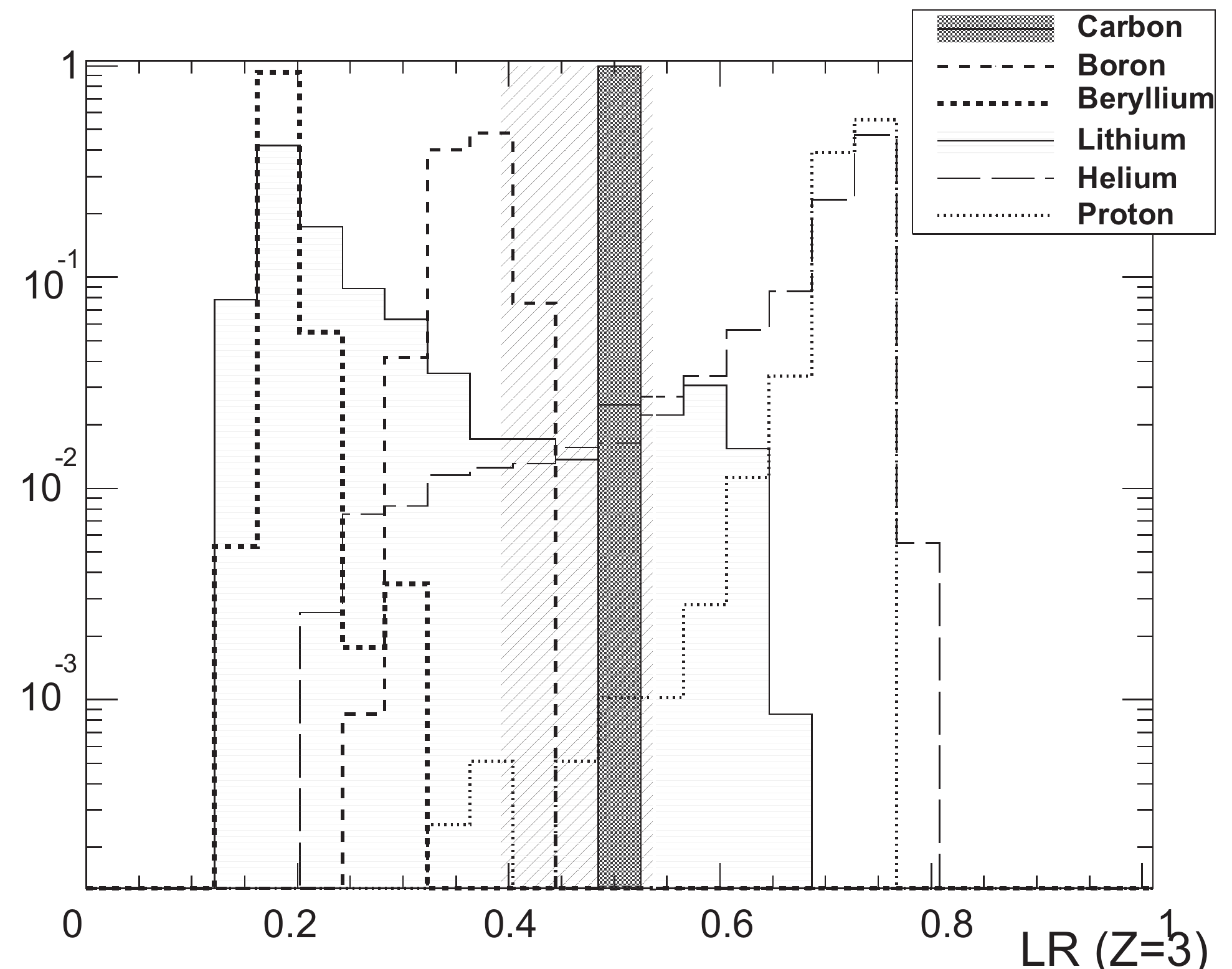}
\caption{\label{fig:likratio}  Likelihood ratio distribution for six
  pure sample of ions: On the left LR is evaluated assuming Z=2, on
  the right assuming Z=3.}
\end{figure}

In Fig.~\ref{fig:likratio} we plotted LR(Z) and LR(Z+1) for the six
“pure samples”, identified by different gray tones, under two Z
hypothesis, namely Z=2 and Z=3. Given our definition of LR, we expect
to observe both a low LR(Z) and a high LR(Z+1) for the sample that
really corresponds to the hypothesis. In fact, we see that the
population whose Z is the hypothesis shows to the left of the value
0.5 in both plots (Z=2 line on the left plot, Z=3 line on the right
plot) so accepting those ions with LR(Z=2) $<$ c1, with c1 $<$ 0.5, we
are selecting the Helium with contamination from Li and Be, however,
asking that LR(Z=3) $>$ c2, with c2 $>$ 0.5 we reject the Li and Be
population (right plot Fig.~\ref{fig:likratio}). The actual test
becomes a series of comparisons, where the event is assigned to one
value of Z if

\begin{equation}
LR(Z)<c_1;  c_1\le 0.48~~~~~~~~LR(Z+1)>c_2;  c_2\ge 0.53\\
\end{equation}

The values for c1 and c2 were chosen looking at the “pure samples”
from a set of data, then we used the same values for all other
targets.

The selection efficiency of the criterion, as well as the
contamination from misenterpreted ions were estimated. To this
purpose, we defined the efficiency and contamination as

\begin{equation}
Eff_Z=\frac{\int_{X_{c1}(Z)}^{X_{c2}(Z)}{f_Z(x)dx}}{\int_{-\infty}^{+\infty}{f_z(x)dx}}
\end{equation}

\begin{equation}
Cont_{Z,C}=\frac{\int_{X_{c1}(Z)}^{X_{c2}(Z)}{f_C(x)dx}}{\sum_k\int{f_K(x)dx}}
\end{equation}

where the integrals at numerator are over the charge domain that would
be selected by our acceptance criterion for a specific value of Z. The
efficiency is the fraction of ions of charge Z that are accepted by
the selection criterion, the contamination is defined as the fraction
of particles identified as charge Z that is statistically expected to
be misinterpreted particle of charge C.

\section{Total charge-changing cross-section and partial fragmentation cross-section}
After the described analysis procedure we obtain a set of numbers
$N(Z)$ that represent the quantity of events where the highest charge
in the event were identified as one specific specie of atomic number
$Z$.  Given the $N(Z)$ quantities, we define
\begin{equation}
P(Z)=\frac{N(Z)}{\sum N(Z)}
\end{equation}
Data taken without a target are used to determine the background for
each fragment Z. We define
\begin{equation}
P_{0}(Z)=\frac{N_{0}(Z)}{\sum N_{0}(Z)}
\end{equation}
which refers to the results obtained with air only.  

The total charge-changing cross-section for a given target of depth
$d$ can be written
\begin{equation}
\sigma_{cc}=-\frac{Alog(P_{corr}(Z_p))}{\rho dN_A}
\end{equation}
where $N_A$ is the Avogrado number, $\rho$ the target density, $A$ the
target mass number, $Z_p$ the atomic number of the primary 
ions from the beam and
\begin{equation}
P_{corr}(Z_p)=\frac{P(Z_p)}{P_{0}(Z_p)}
\end{equation}

The error on $\sigma_{cc}$ is given by
\begin{equation}
\frac{\delta \sigma_{cc}}{\sigma_{cc}}=-\frac{\delta P_{corr}(Z_p)}{P_{corr}(Z_p)log(P_{corr}(Z_p))}
\end{equation}

Partial fragment production cross-sections are given by
\begin{equation}
\sigma_Z=\sigma_{cc}\frac{P_{corr}(Z)}{1-P_{corr}(Z_p)}
\label{eq:partial}
\end{equation}

In order to derive the cross-section results we considered that a
small fraction of the produced fragments can be out of geometrical
acceptance of the detector, especially for low Z fragments which have
the highest transverse momentum.

\begin{figure}
\centering
\includegraphics[width=3.5in]{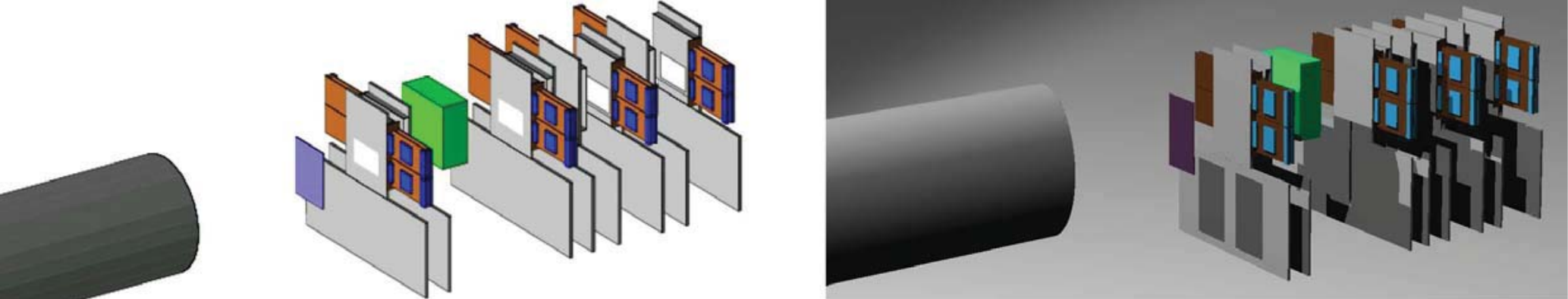}
\caption{\label{fig:gsisim} Visualisation in GEANT4 (left) and FLUKA
  (right) of the geometrical modeling of experimental setup used in
  this work.}
\end{figure}

In addition, in any target, there is a finite probability for
secondary interactions involving fragments. While these have no effect
on the determination of the total cross-section, they affect fragment
yield enhancing the number of lighter fragments.  In order to estimate
both the effects of geometrical acceptance and the fragment
reinteractions Monte Carlo simulation of the full experimental setup
were performed with GEANT4 and FLUKA.  
For this purpose we have
developed a PYTHON based simulation framework that uses FLUKA and/or
GEANT4 for the run of the Monte Carlo simulations and offer a common
set of simple functions for the definition of the beam, of the
geometry, and for the post-processing analysis.  For the GEANT4
simulations the geometry of the different experimental setup were
defined in form of Geometry Data Markup Language (GDML) files.  These
GDML files were translated into FLUKA geometry input cards by a PYTHON
GDML-to-FLUKA geometry translator developed by
us. Fig.~\ref{fig:gsisim} represents the mo\-de\-ling of one of the
experimental set-up considered in this project as seen in GEANT4
(left) and in FLUKA (right) visualisation drivers. The PYTHON interface we have developed was to avoid any error which may have been introduced by entering into the FLUKA the parameters of a complex geometry, as in this work, via data cards. The interface takes simply the parameters set from GDML description and converts it to FLUKA data card readable values. Therefore, existing codes such as  MRED   \cite{mred} and others  \cite{tiara, musca}  more complicated than ours and they point to solve higher order problems.

Large Monte Carlo samples of different experimental configurations
tested at GSI were produced and analysed. The geometrical
inefficiencies were then evaluated and the cross-section values were
corrected accordingly. The cross-sections determined for several
thicknesses and of same material and for several experimental
configurations have been combined to enhance the statistical accuracy
of the measurements.
 
The error on total cross-section includes both statistical and
systematic contributions. Main contributions to systematic error
originate from two effects: one is the difference in cross-sections
obtained by using different target thickness of the same material and
different experimental configurations, the other is the geometrical
acceptance calculations obtained by Monte Carlo runs using different
physics lists produce different fragment angular distributions (see
Fig.~\ref{fig:angWater}). The overall contribution gives a
systematic uncertainty of about $3\%$.

The fragment production cross-section is proportional to the total
charge-changing cross-section (eq.~\ref{eq:partial}), therefore any
systematic error in the latter propagates to systematic error in the
former.  In addition systematic errors due to ambiguous cases where
there is an inefficiency in charge assignment procedure have been
evaluated and accounted for.

The summary of the measurements on all tested target materials,together 
with literature data, are reported in Table III. 
In Table IV the partial fragment production ratio with 
respect to the total charge-changing cross-section for all tested 
target materials are listed.

\begin{table*}[!ht]
\label{tab:totcross}
\begin{center}
\caption{Total charge-changing cross-section: comparison with previous
  measurements. The numbers in the fourth column are the relative
  errors (in \%) referring to calculated values for this experiment or to data from literature.}
\begin{tabular}{|c|c|c|c|c|}
\hline
 Energy (MeV/n) & Target & Cross-Section (mbarn) & Rel. Error (\%) & Ref.\\
\hline
200 & graphite & 658 (7) & 1.1 &\cite{weber} \\
\hline
267 & graphite & 748 (19) & 2.5 &\cite{schall} \\
\hline
290 & graphite & 706 (7) & 1.0 &\cite{cecchini}
\\
\hline
400 & graphite & 672 (7) & 1.0 &\cite{weber}\\
\hline
400 & graphite & 713 (11) & 1.5 &\cite{zeitlin}\\
\hline
498 & graphite & 758 (15) & 2.0 &\cite{schall}\\
\hline
500 & graphite & 703 (18) & 2.5 &This exp.\\
\hline
\hline
192 & aluminium & 1179 (29) & 2.5 &\cite{schall}\\
\hline
267 & aluminium & 1078 (17) & 1.6 &\cite{schall}\\
\hline
290 & aluminium & 1155 (108) & 9.3 &\cite{cecchini}\\
\hline
290 & aluminium & 1052 (11) & 1.0 &\cite{zeitlin}\\
\hline
400 & aluminium & 1011 (9) & 0.9 &\cite{zeitlin}\\
\hline
498 & aluminium & 1103 (28) & 2.5  &\cite{schall}\\
\hline
500 & aluminium & 1095 (26) & 2.5 &This exp.\\
\hline
676 & aluminium & 1096 (100) & 9.1 &\cite{schall}\\
\hline
\hline
500 & iron & 1509 (37) & 2.5 & This exp.\\
\hline
\hline
290 & copper & 1625 (18) & 1.1 &\cite{zeitlin}\\
\hline
400 & copper & 1557 (10) & 0.6 &\cite{zeitlin}\\
\hline
500 & copper & 1598 (50) & 3.1 &This exp.\\
\hline
\hline
290 & tin & 2069 (18) & 0.9 &\cite{zeitlin}\\
\hline
400 & tin & 2035 (21) & 1.0 &\cite{zeitlin}\\
\hline
500 & tin & 2141 (79) & 4.9 & This exp.\\
\hline
\hline
500 & tantalum & 2936 (105) & 3.6 &This exp.\\
\hline
\hline
290 & lead & 2795 (15) & 0.5 &\cite{zeitlin}\\
\hline
400 & lead & 2745 (45) & 1.6 &\cite{zeitlin}\\
\hline
500 & lead & 2926 (116) & 3.9 &This exp.\\
\hline
\hline
192 & water & 1264 (16) & 1.3 &\cite{schall}\\
\hline
267 & water & 1163 (13) & 1.1 &\cite{schall}\\
\hline
326 & water & 1250 (51) & 4.1 &\cite{toshito}\\
\hline
352 & water & 1202 (47) & 3.9 &\cite{toshito} \\
\hline
377 & water & 1253 (45) & 3.6 &\cite{toshito} \\
\hline
498 & water & 1220 (20) & 1.6 &\cite{schall} \\
\hline
500 & water & 1211 (27) & 2.2 & This exp.\\
\hline
670 & water & 1261 (13) & 1.0 &\cite{schall}\\
\hline
\hline
192 & lucite & 7250 (102) &  1.4 &\cite{schall}\\
\hline
267 & lucite & 6733 (74) & 1.1 &\cite{schall}\\
\hline
464 & lucite & 7019 (112) & 1.6 &\cite{schall}\\
\hline
500 & lucite & 7051 (165) & 2.3 &This exp.\\
\hline
676 & lucite & 7170 (360) & 5.0 &\cite{schall}\\
\hline
\hline
192 & polyethylene & 1157 (13) & 1.1 &\cite{schall}\\
\hline
267 & polyethylene & 1075 (11) & 1.0 &\cite{schall}\\
\hline
498 & polyethylene & 1135 (15) & 1.3 &\cite{schall}\\
\hline
500 & polyethylene & 1120 (54) & 7.0 &This exp.\\
\hline
\hline
500 & silicon carb. & 1974 (91) & 4.9 &This exp.\\
\hline
\end{tabular}
\end{center}
\end{table*}

\begin{table*}[!ht]
\label{tab:partialcross}
\begin{center}
\caption{Ratio of partial fragmentation cross-section over total
  charge-changing cross-section for $^{12}C$ ion at 500 MeV/n energy
  for all tested target materials.}
\begin{tabular}{|c|c|c|c|c|c|}
\hline
\hline
 Target & $Z=5$ & $Z=4$ & $Z=3$ & $Z=2$ & $Z=1$\\
\hline
\hline
graphite & 10.9 (0.6) & 6.3 (0.4) & 11.9 (1.5) & 56.9 (2.8) & 14.1 (1.4)\\
\hline
aluminium & 9.7 (0.5) & 5.0 (0.3) & 9.2 (1.2) & 55.7 (3.4) & 20.3 (1.9)\\
\hline
iron & 8.3 (0.4) & 4.2 (0.3) & 10.0 (1.2) & 52.2 (3.5) & 25.2 (2.1)\\
\hline
copper & 8.7 (0.8) & 5.8 (0.6) & 8.7 (2.2) & 50.8 (5.8) & 26.0 (4.4)\\
\hline
tin & 7.1 (0.6) & 4.5 (0.5) & 7.9 (1.4) & 50.3 (4.8) & 30.2 (3.3)\\
\hline
tantalum & 6.5 (0.5) & 3.8 (0.4) & 7.3 (0.9) & 49.3 (3.2) & 33.1 (2.4)\\
\hline
lead & 7.0 (0.7) & 4.0 (0.5) & 6.6 (1.2) & 48.6 (4.2) & 33.9 (3.2)\\
\hline
water & 13.9 (0.5) & 6.1 (0.3) & 13.3 (2.0) & 56.4 (4.0) & 10.4 (1.7)\\
\hline
lucite & 13.3 (0.5) & 6.5 (0.3) & 12.1 (2.0) & 56.7 (4.3) & 11.5 (1.5)\\
\hline
polyethylene & 13.3 (1.6) & 5.3 (0.9) & 11.5 (1.9) & 60.4 (5.7) & 9.4 (1.6)\\
\hline
silicon carb. & 9.2 (1.3) & 6.1 (1.0) & 10.2 (2.6) & 57.1 (7.2) & 17.4 (3.1)\\
\hline
\hline
\end{tabular}
\end{center}
\end{table*}

\section{Comparison of results with GEANT4 and FLUKA simulations}
\begin{figure}
\centering
\includegraphics[width=3.5in]{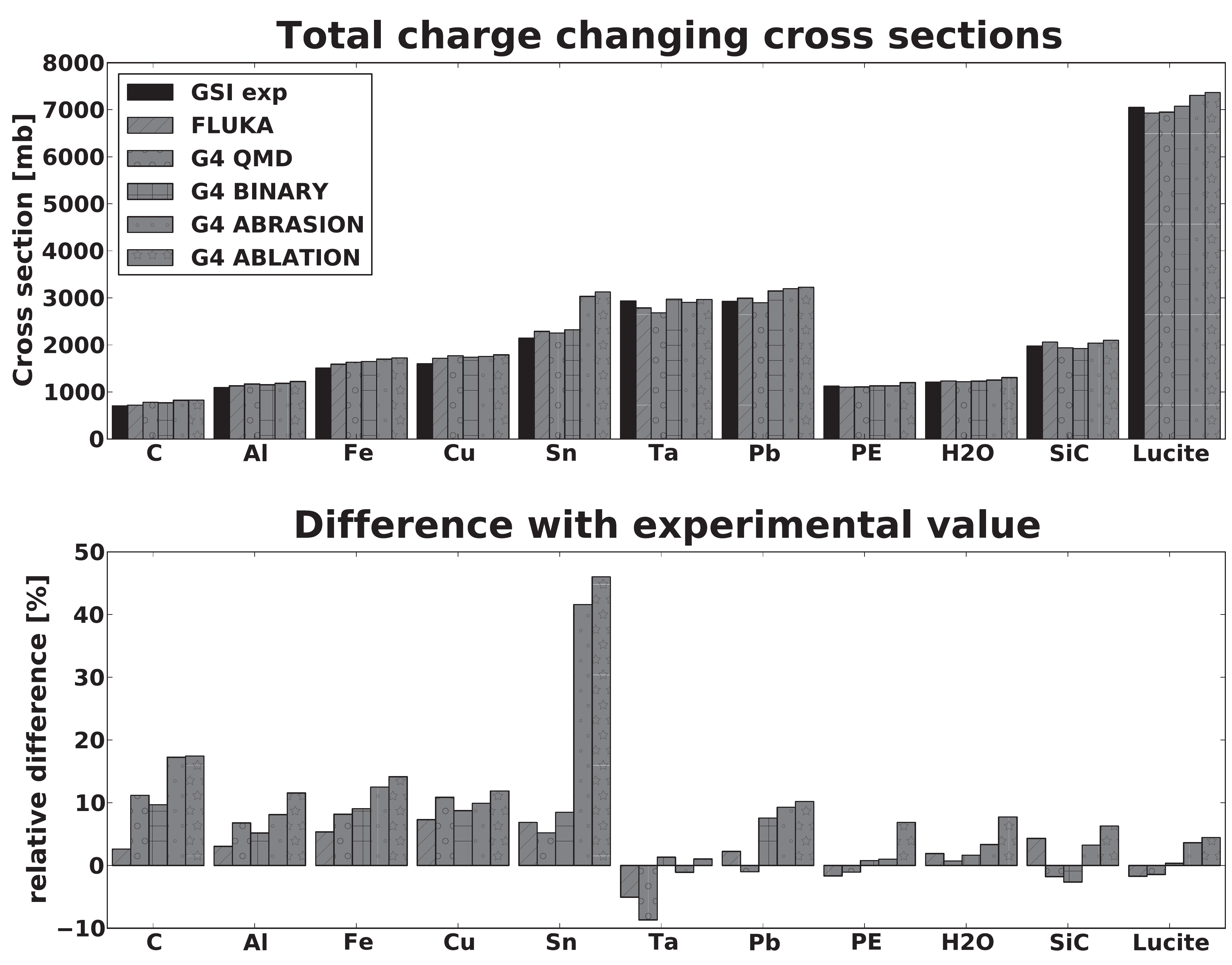}
\caption{\label{fig:totalCS} Total charge-changing cross-section for
  500 MeV/n $^{12}C$ ion as a function of target type, 
compared with different physics models of GEANT4 and FLUKA. The bottom panel
  represents the relative differences between the model
  and the experimental cross-section.}
\end{figure}

The measured total charge-changing cross-sections has been compared
with the predictions provided by nuclear ion-ion interaction models
implemented in FLUKA and GEANT4 simulation tools
(Fig.~\ref{fig:totalCS}).  Except for Abrasion and Ablation models
there is an agreement within 10\% between simulation and data for all
targets.  For several targets FLUKA shows an agreement with data
within 3-5\%.

\begin{figure}
\centering
\includegraphics[width=3.5in]{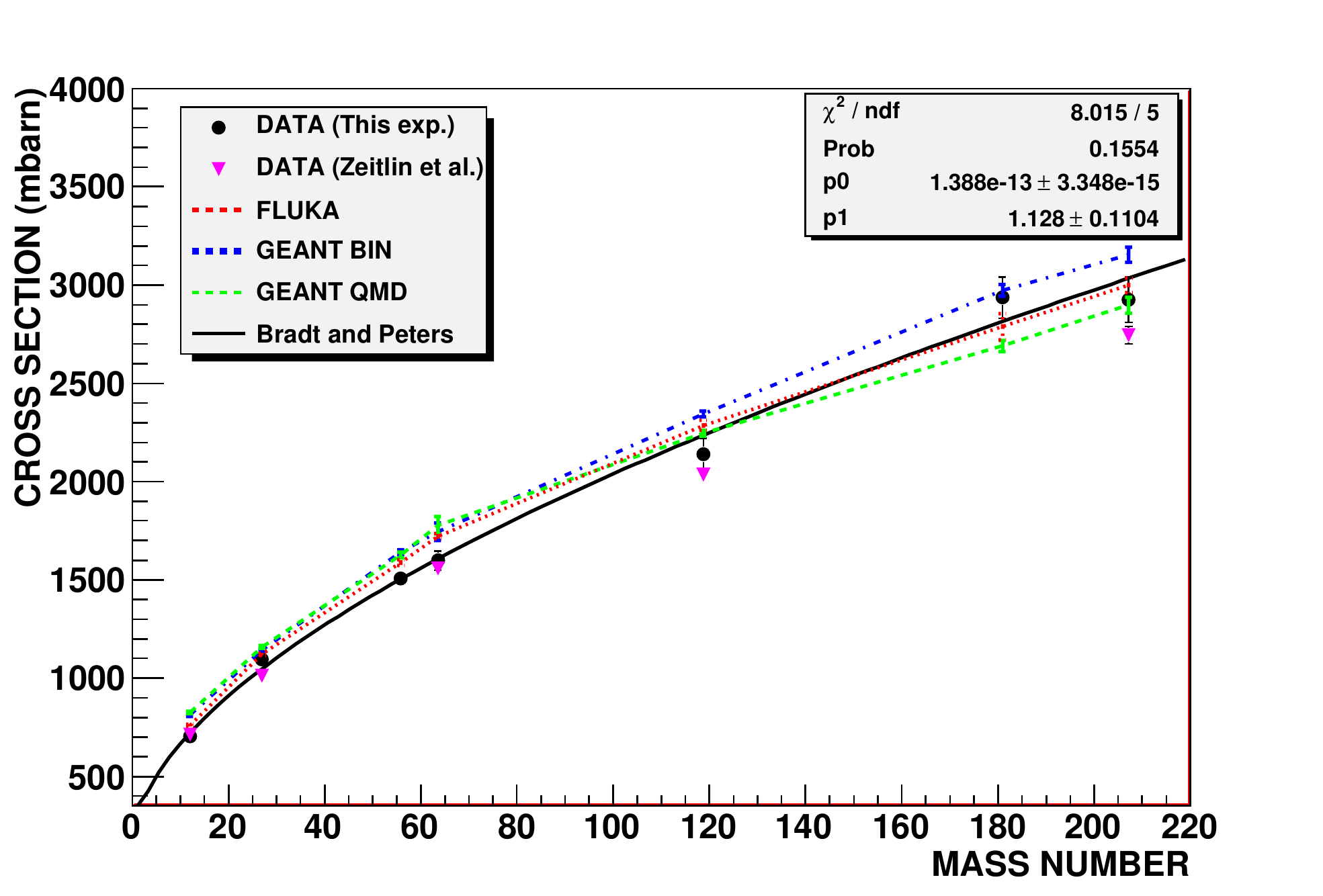}
\caption{\label{fig:crossVStarg} Experimental total charge-changing
  cross-sections measured in this experiment for elemental target as a
  function of target mass number $A$. In the plot model predictions
  from GEANT4 and FLUKA togheter with the measurement given
  in~\cite{zeitlin} are reported. The solid line is the fit to data as
  measured in this work with the Bradt and Peters curve.  }
\end{figure}

\begin{figure*}[!htbp]
\centering
\includegraphics[width=1.74in]{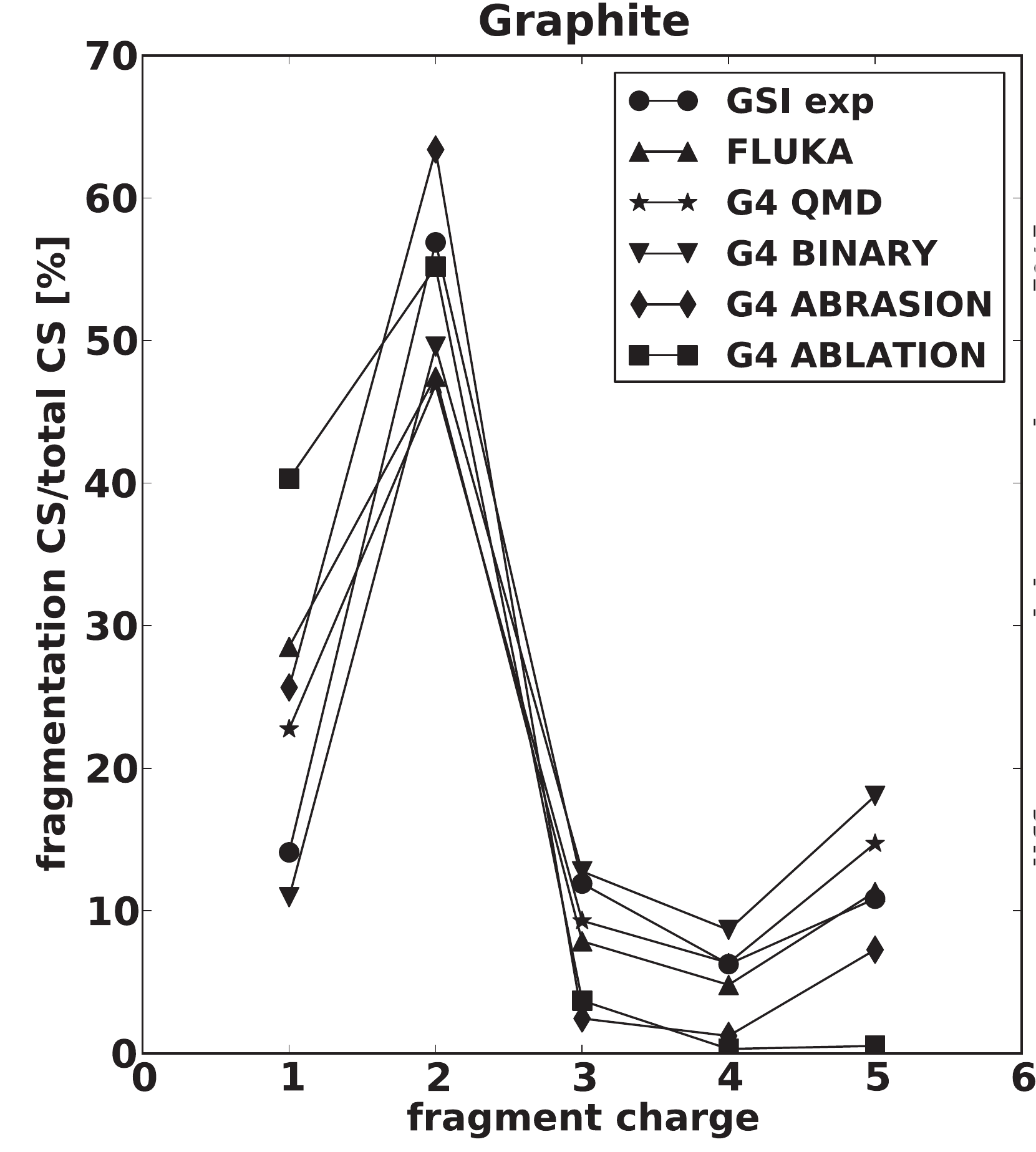}
\includegraphics[width=1.74in]{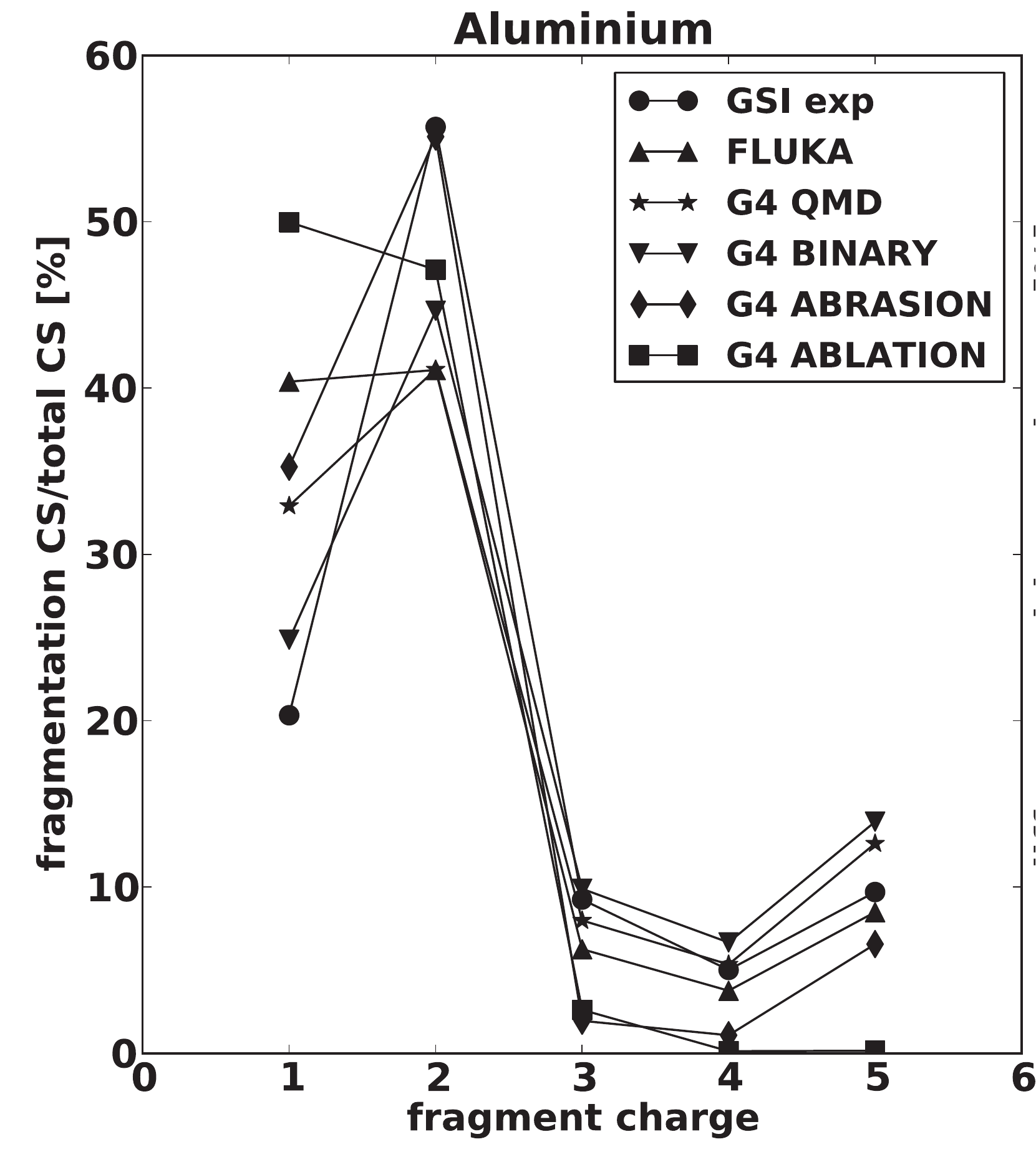}
\includegraphics[width=1.74in]{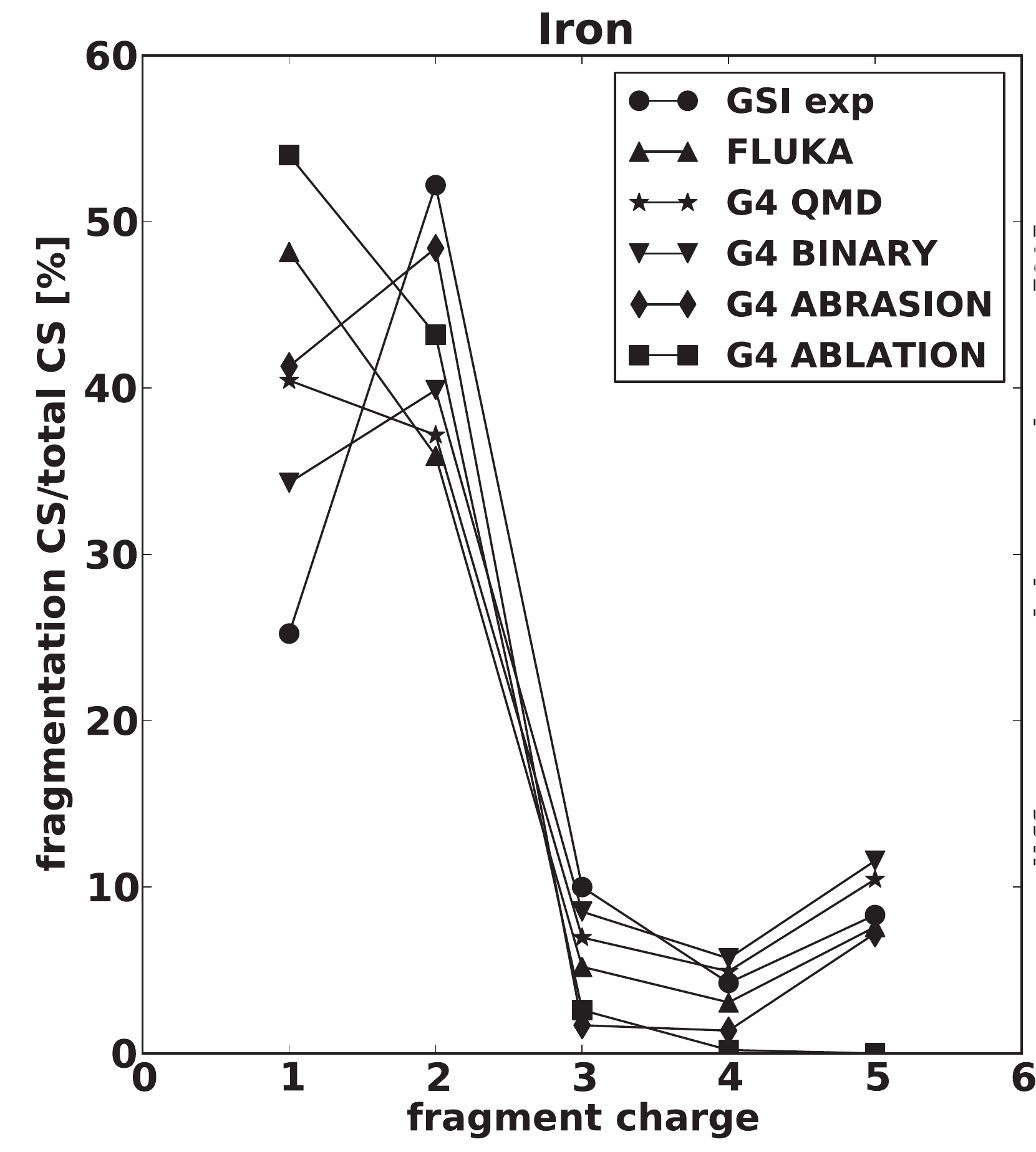}
\includegraphics[width=1.74in]{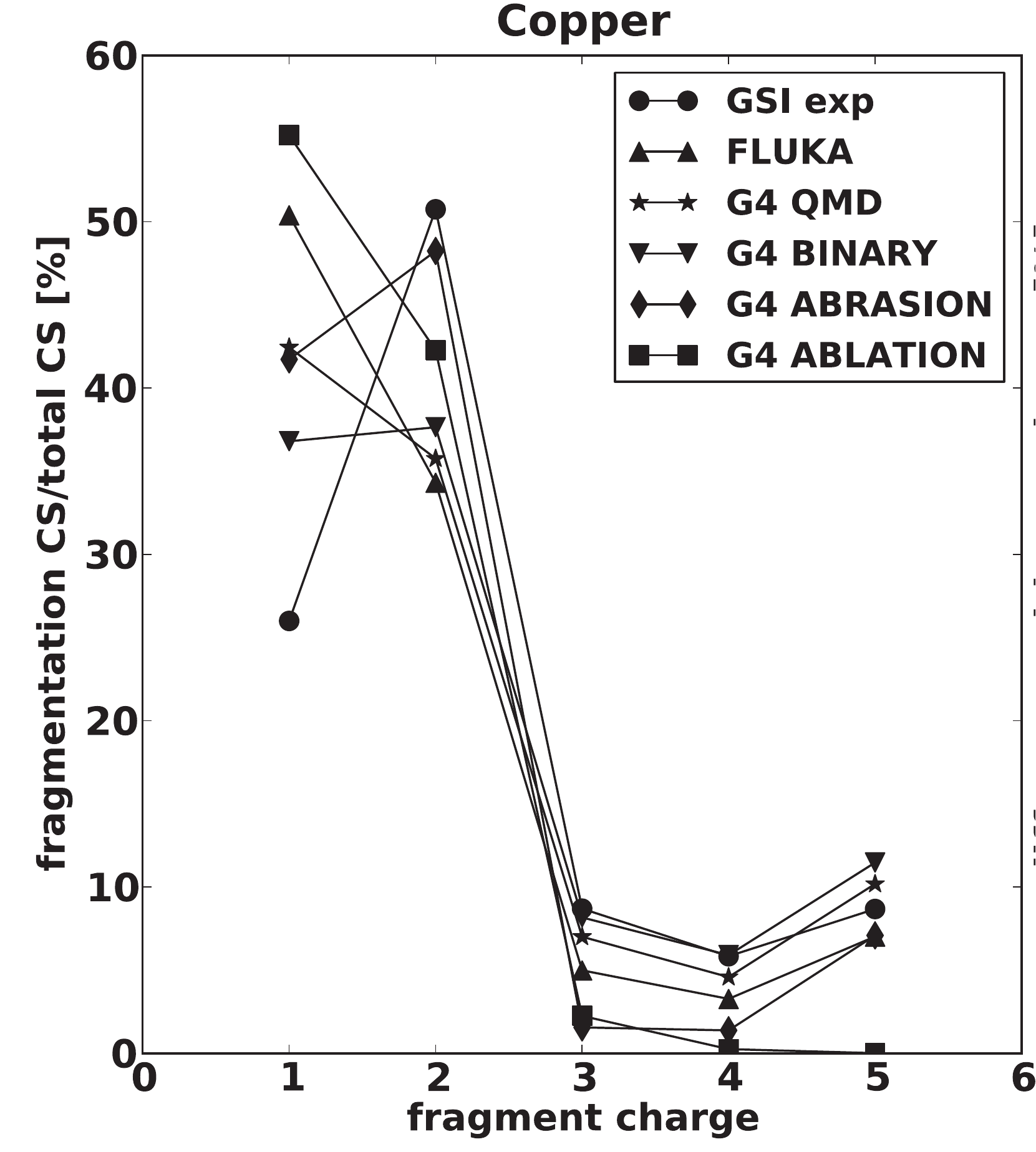}
\includegraphics[width=1.74in]{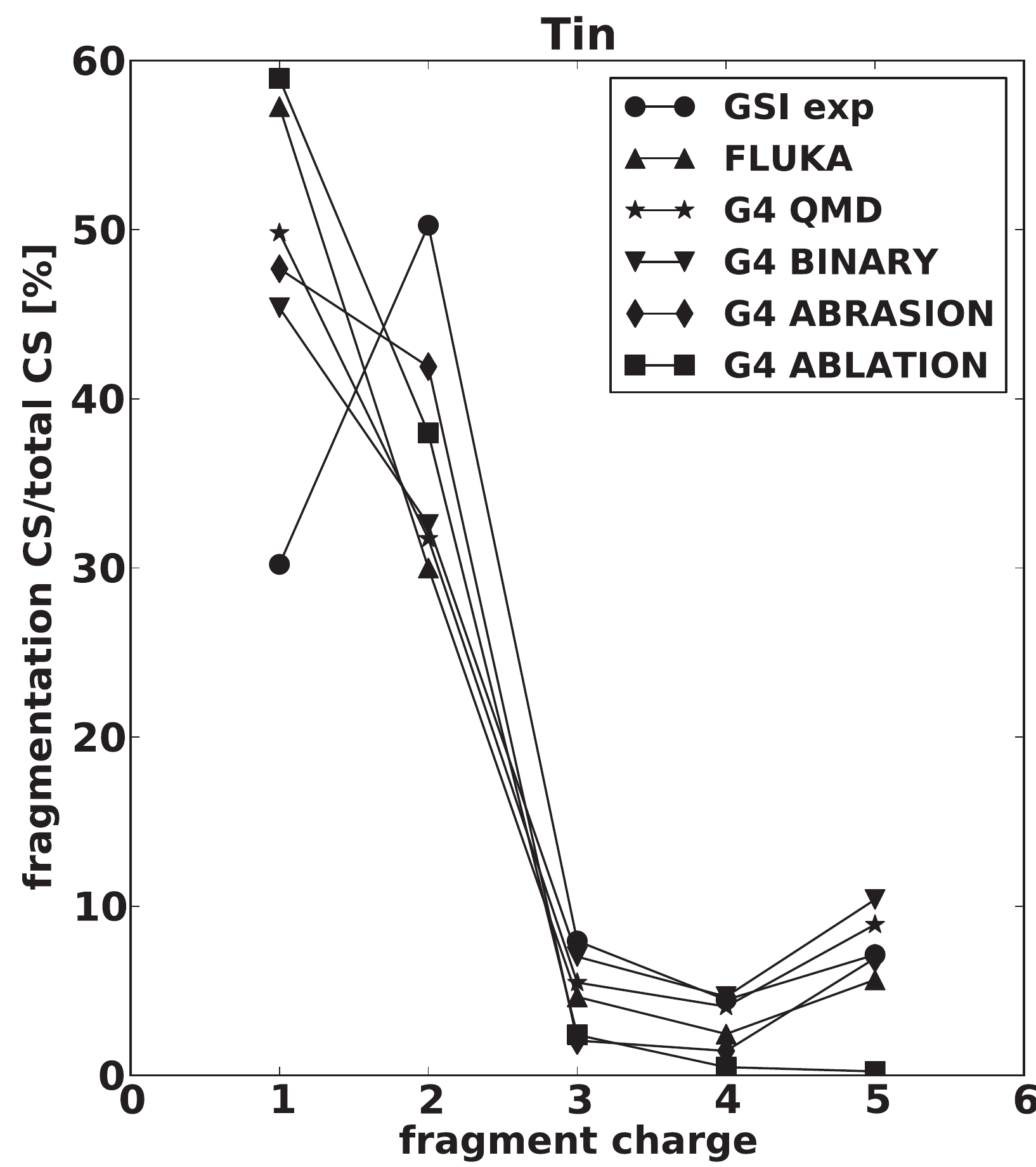}
\includegraphics[width=1.74in]{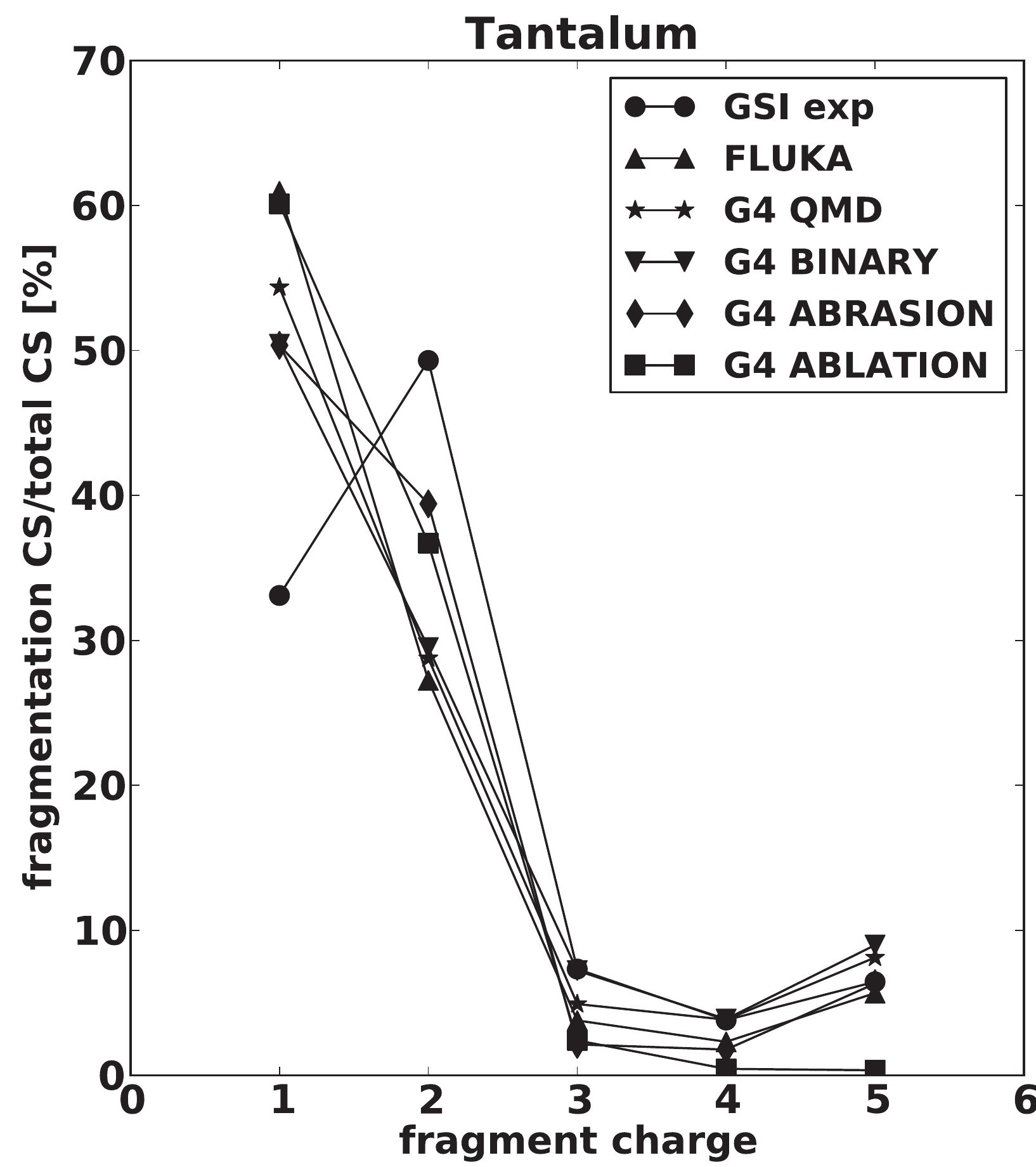}
\includegraphics[width=1.74in]{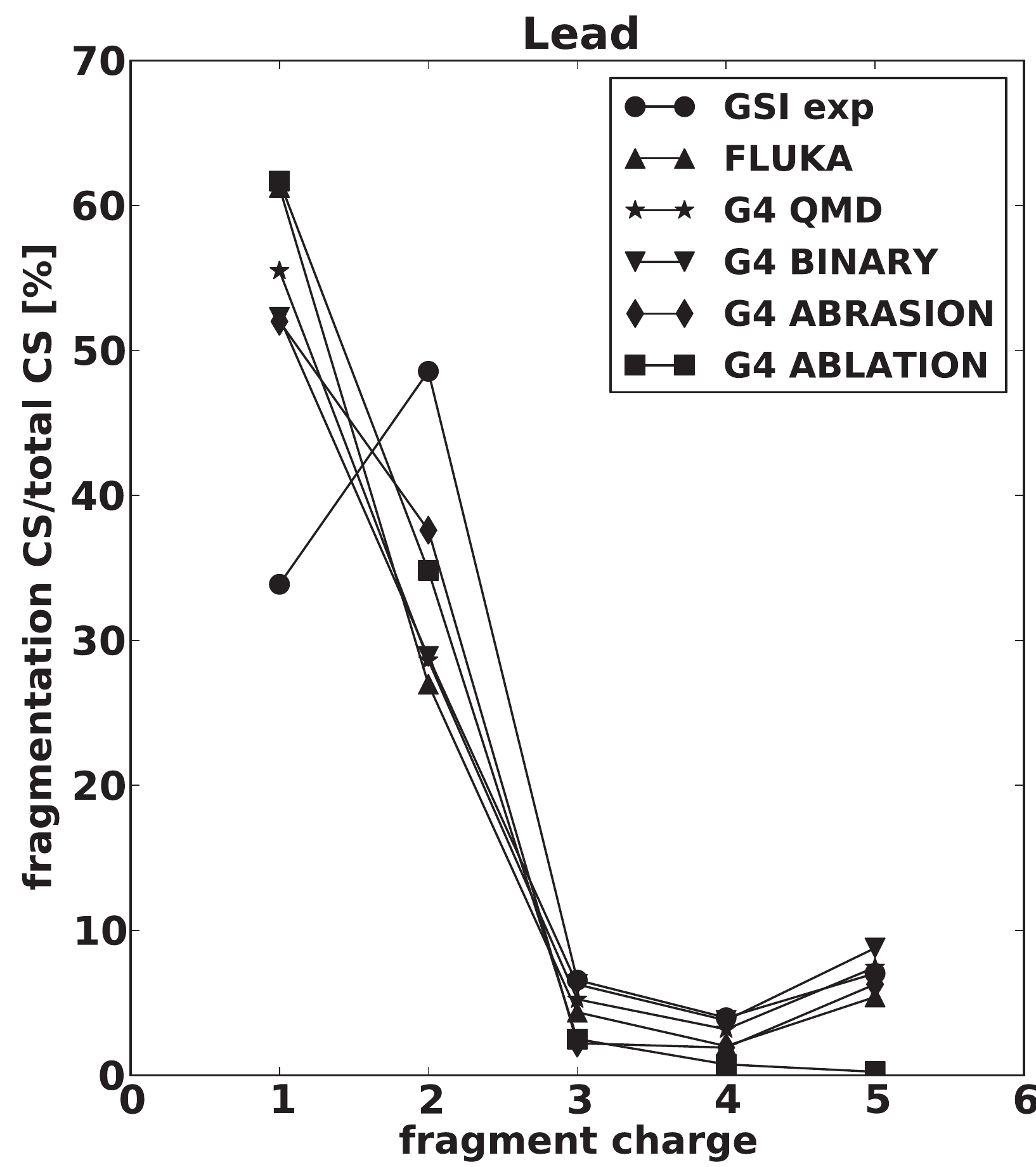}
\includegraphics[width=1.74in]{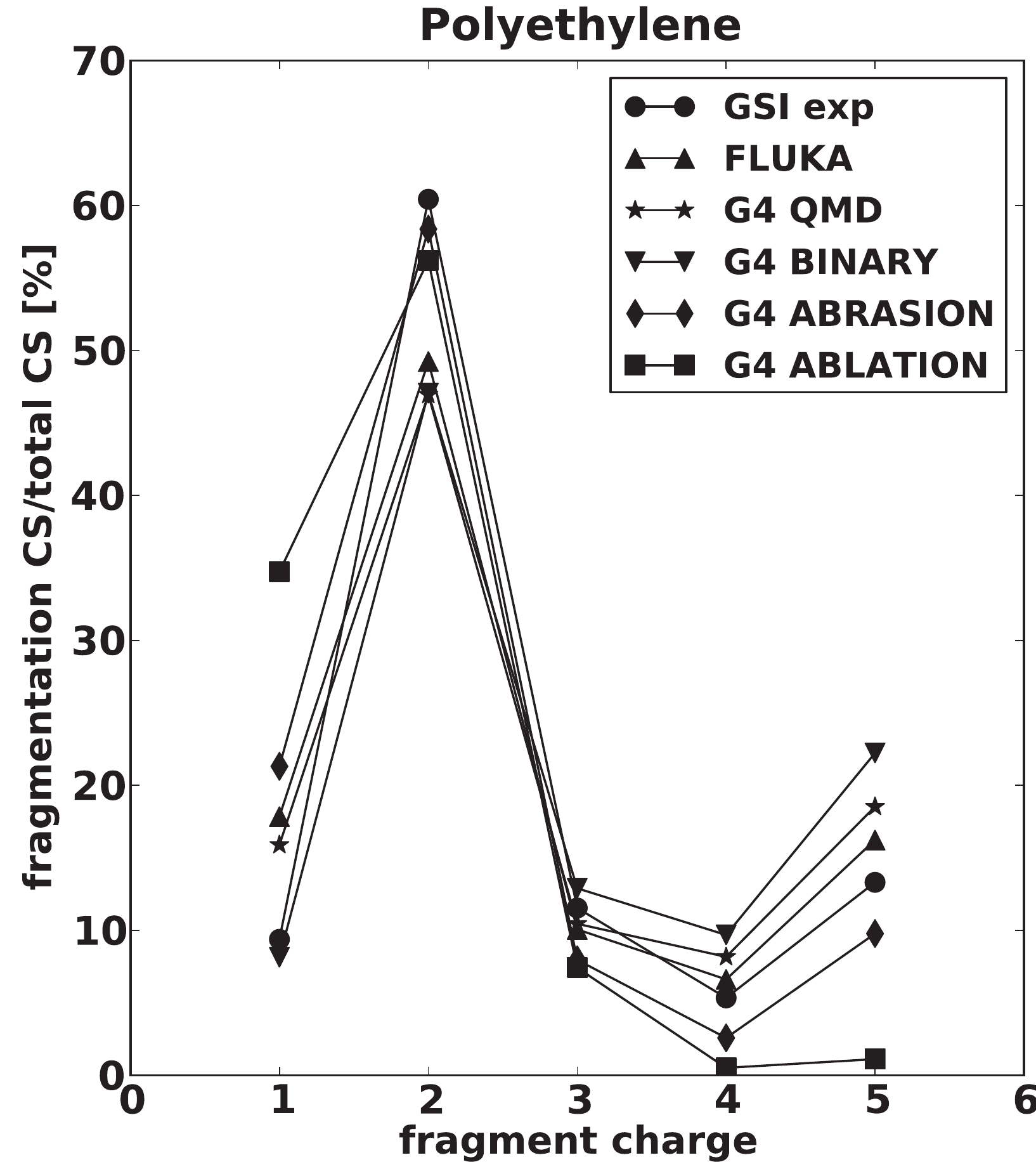}
\includegraphics[width=1.74in]{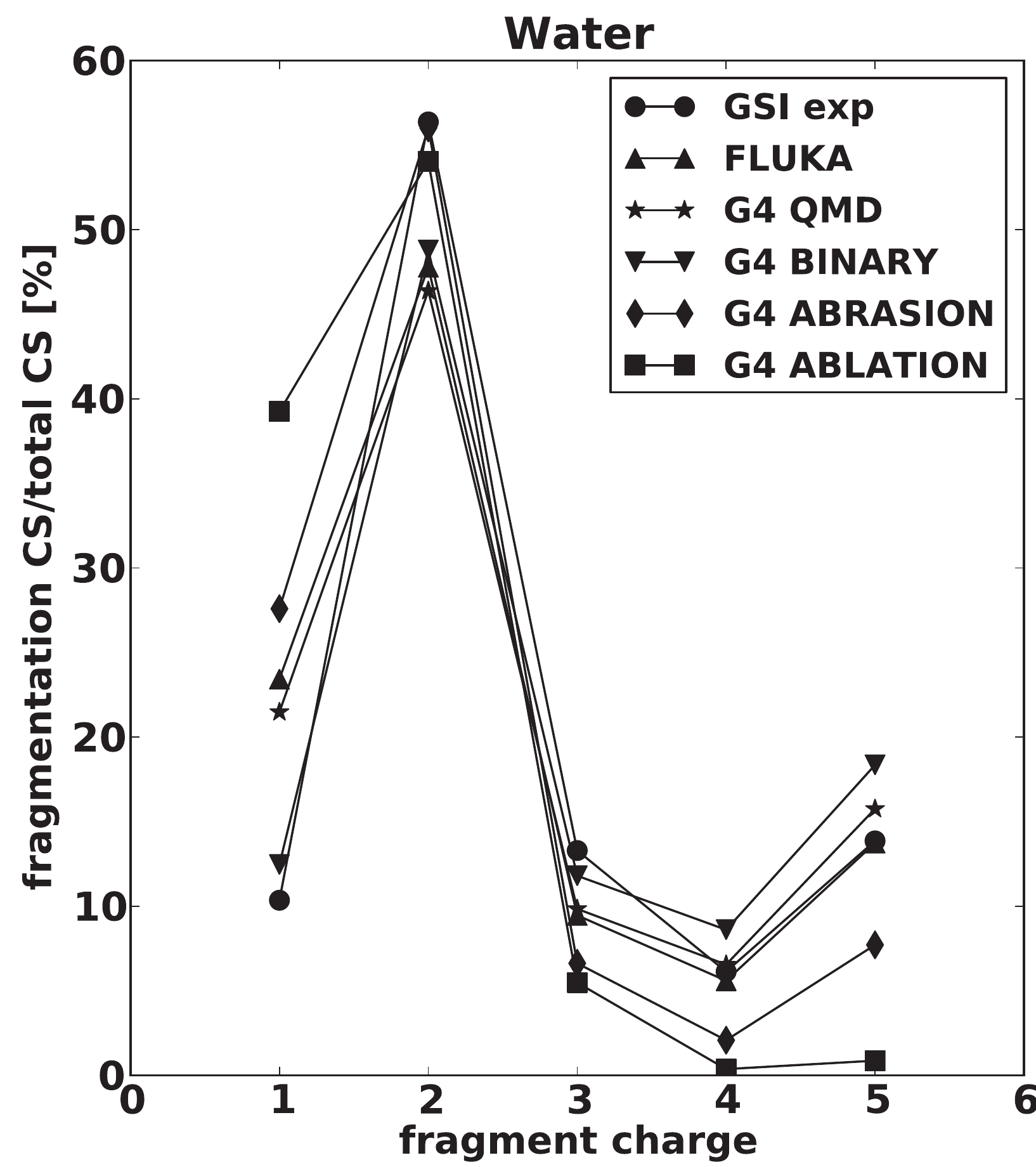}
\includegraphics[width=1.74in]{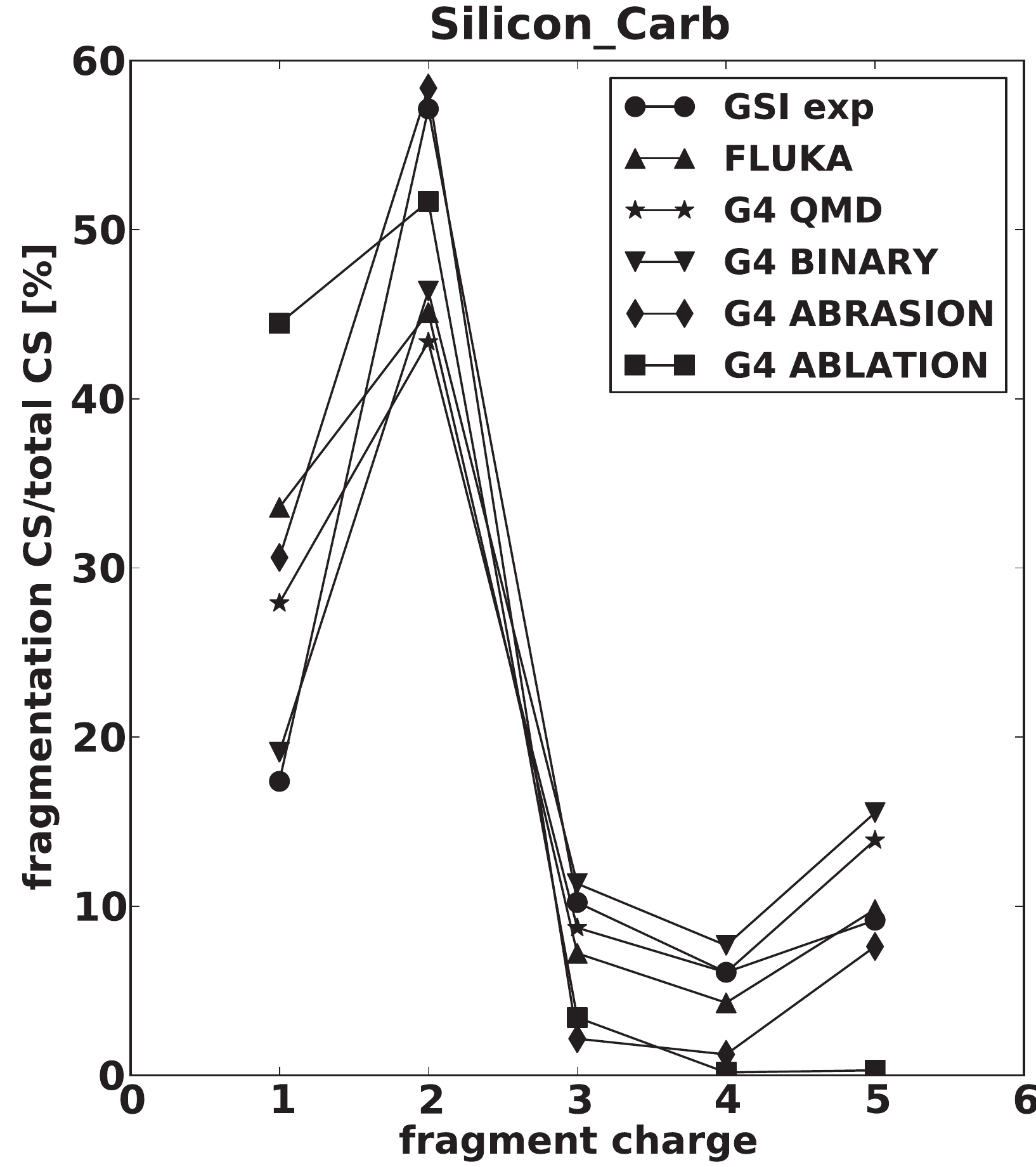}
\includegraphics[width=1.74in]{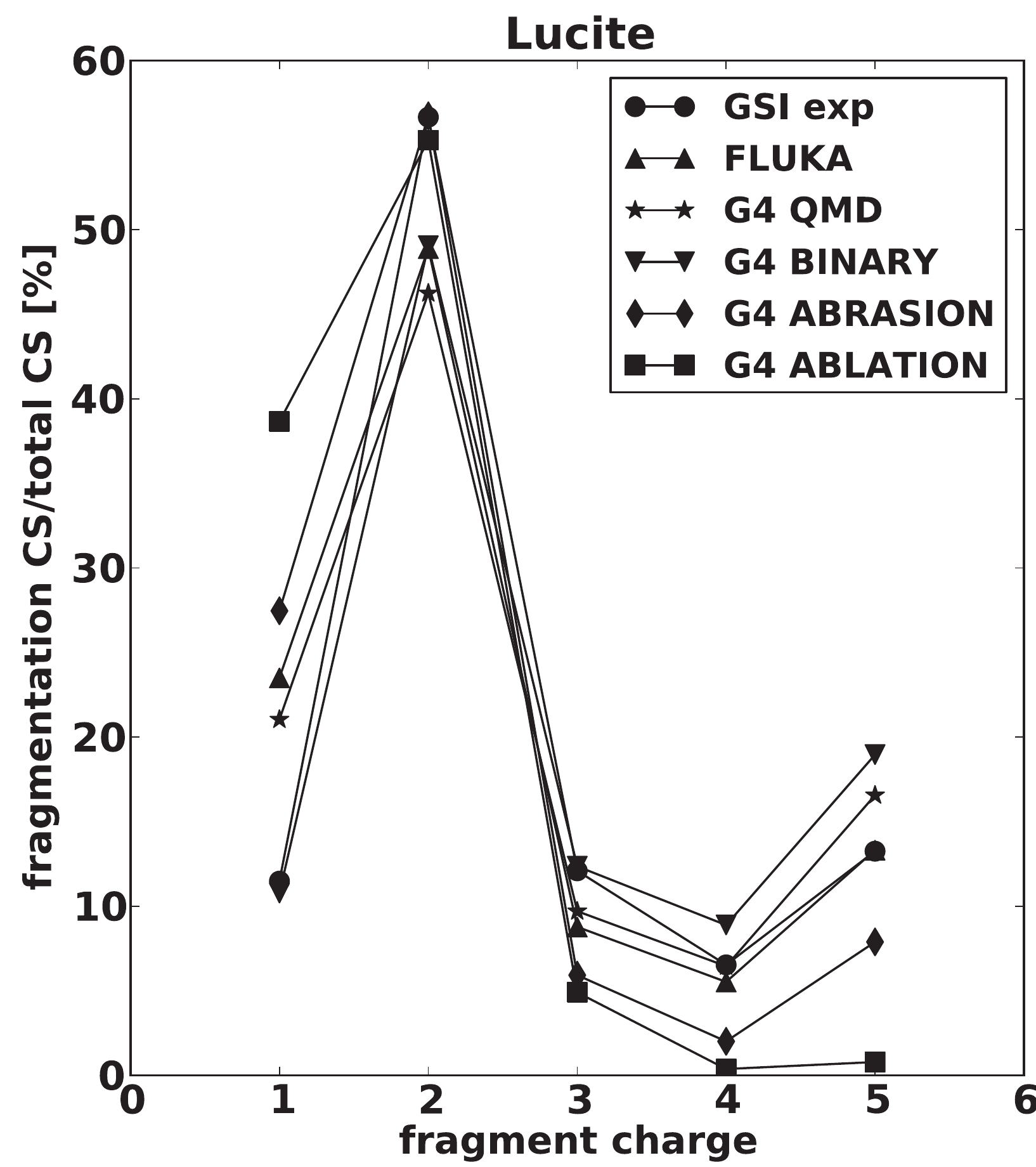}%
\caption{\label{fig:partial} Measured ratio of fragmentation
  cross-section over total charge-changing cross-sections for $^{12}C$
  ion at the energy of 500 MeV/n interacting with different targets
  compared with different physics models of GEANT4 and FLUKA. }
\end{figure*}

Fig.~\ref{fig:crossVStarg} shows the variation of total
charge-changing cross-sections as a function of target mass number $A$
for elemental targets.
Data have been fitted with the simple geometric cross-section model
calculation based on Bradt and Peters \cite{bradtAndPet} parametrisation

\begin{equation}
\sigma(A_p,A_t)=\pi r_0^2\big (A_p^{1/3}+A_t^{1/3}-b \big )^2
\end{equation}

where $A_p$ and $A_t$ are the mass number of the projectile and
target, respectively.
The results of the fit gives the values $r_0=1.39\pm0.03$ and
$b=1.13\pm0.11$ to be compared with the values of $r_0=1.29$ and
$b=0.86$ as in~\cite{zeitlin} and $r_0=1.35$ and $b=0.83$ as
in~\cite{chen}.

Fig.~\ref{fig:partial} shows the ratio of fragmentation
cross-sections over the total charge-changing cross-sections for
different target materials, for GEANT4 models, FLUKA and experimental
data.  On the x axis the fragment charge refers to the charge number Z
of the heaviest fragment. For Z$\geq$3 the G4QMD, G4 Binary and FLUKA
models agree reasonably well with this experiment. For $Z=2$ (only He
and H isotopes are produced) and Z=1 (only H isotope is produced), the
Monte Carlo models show larger discrepancies compared to the
experimental data. In general the models under-estimate the
fragmentation cross-section for Z=2 and over-estimate it for Z=1. In
addition the discrepancy is higher for high mass number targets with
respect to low mass number targets.

\begin{figure*}[!htbp]
\centering{
\includegraphics[width=2.8in]{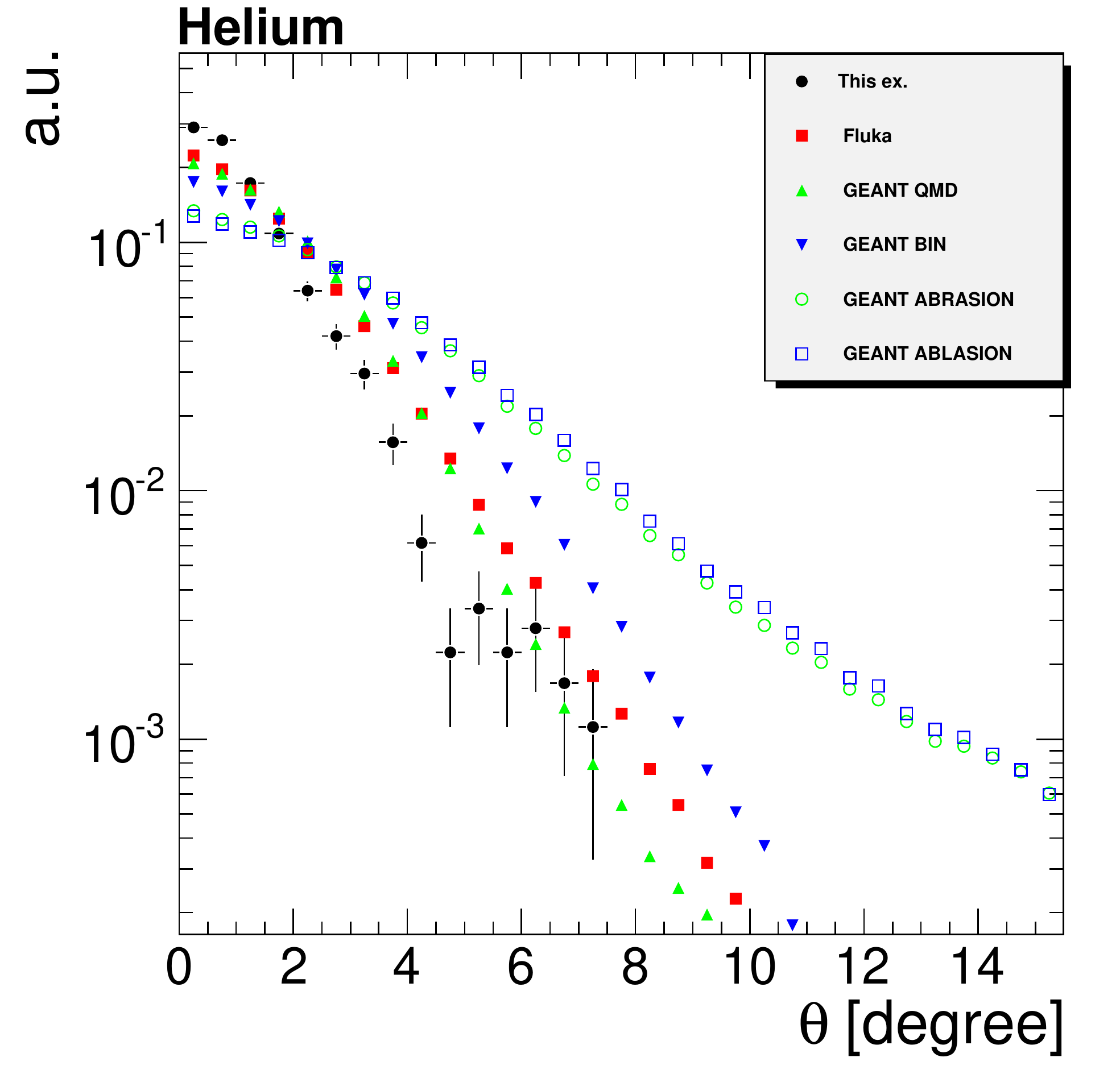}
\includegraphics[width=2.8in]{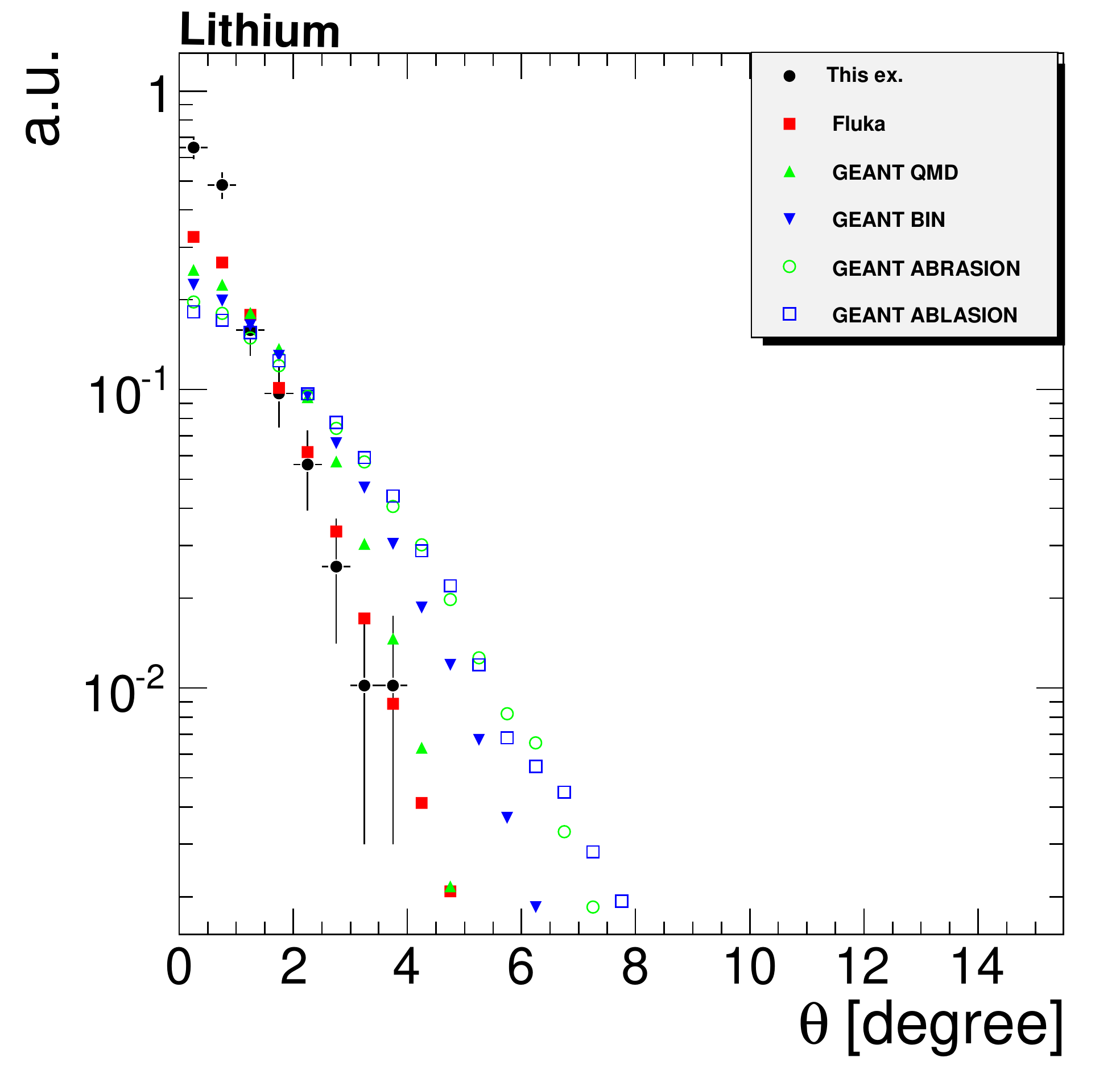}
\includegraphics[width=2.8in]{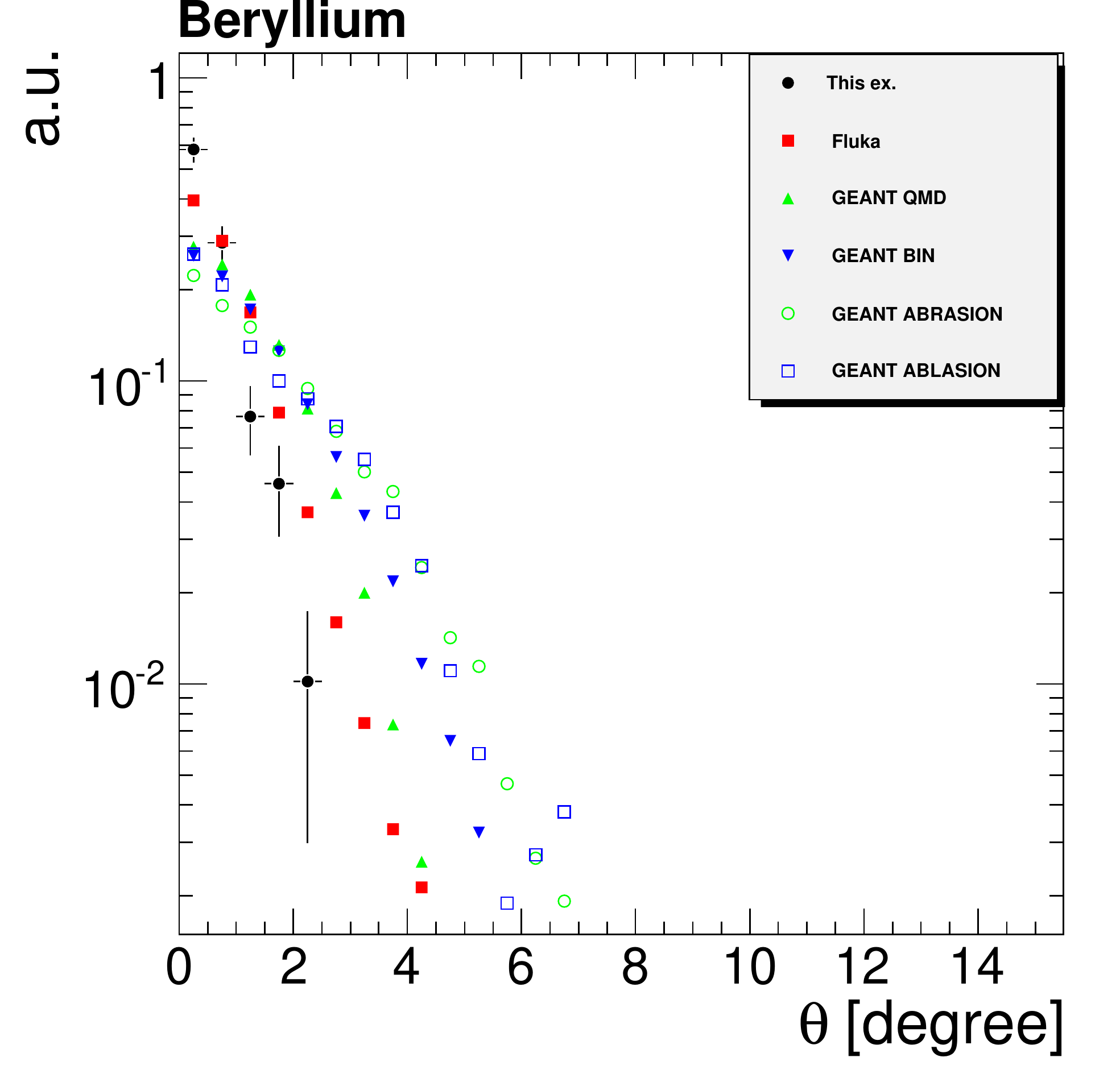}
\includegraphics[width=2.8in]{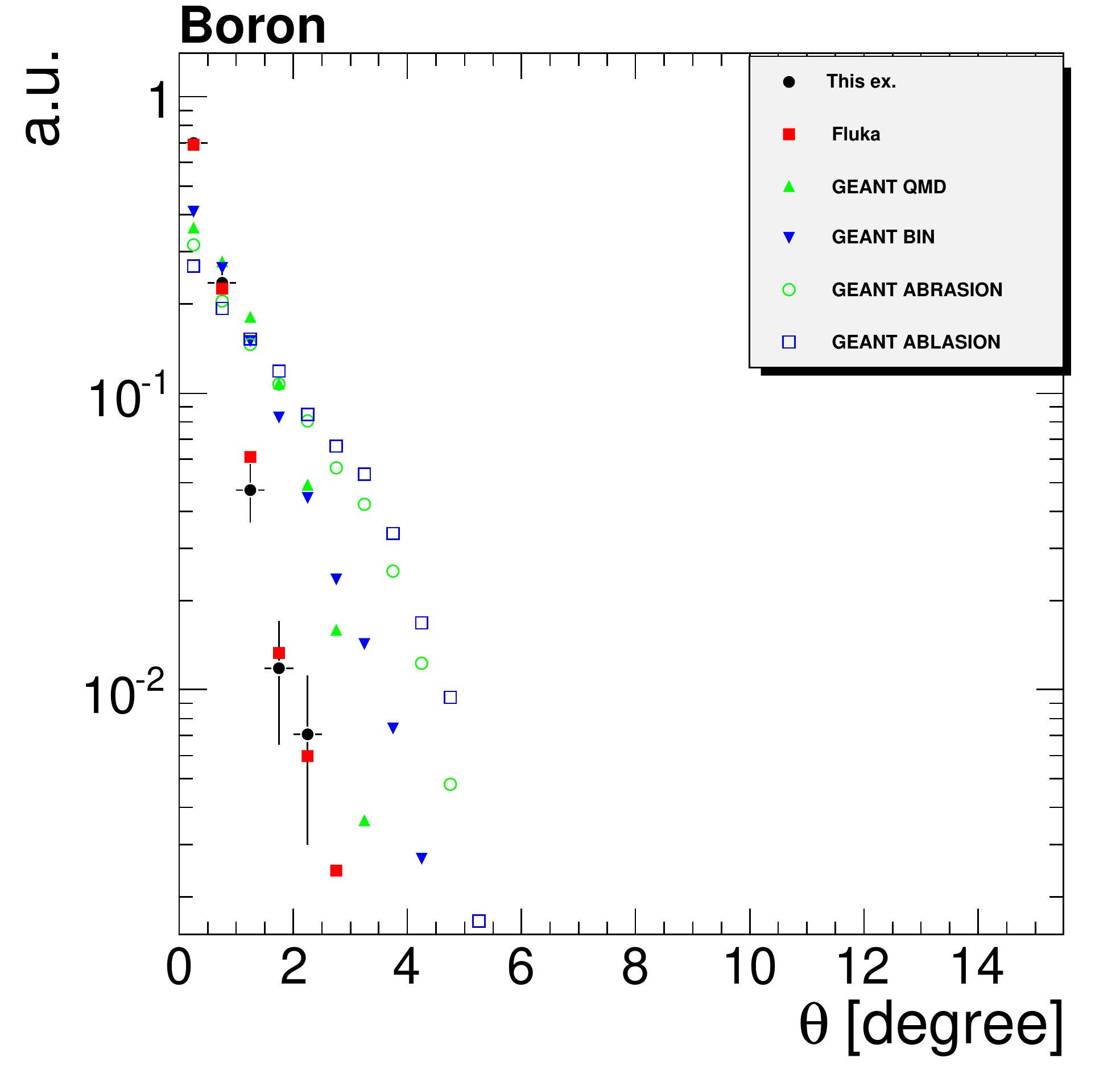}
\caption{\label{fig:angWater} Angular distribution of fragments events
  produced in the interaction of 500 MeV/n $^{12}C$ ions on water
  target compared with different physics models of GEANT4 and FLUKA:
  Helium (top left), Lithium (top right), Beryllium (bottom left),
  Boron (bottom right).}
}
\end{figure*}

From the track fitting of the experimental data the angle $\Theta$
between the beam-line direction and the XZ projection of the direction
of the highest-Z fragment after the target, can be
deduced. In Fig.~\ref{fig:angWater} the angular distribution is shown
for fragment events in water target and compared with Monte Carlo
models.

\section{Conclusions}
We measured at GSI, Darmstadt, the total charge-changing and partial fragment 
cross-sections for the interactions of $^{12}C$ at 500
MeV/n on several targets.  

A PYTHON based simulation framework has been developed to simulate
with GEANT4 and FLUKA the interaction of the $^{12}C$ beam with the
different experimental setups considered in this study. The comparison
of measured data with the simulation models gives a reasonable
agreement for total charge-changing cross-sections.  The detailed
analysis of $^{12}C$ data has shown that there is a discrepancy
between Monte Carlo model predictions and data on low Z (Z=1 and Z=2)
cross-section values for almost all target materials. The discrepancy
is systematic such that the data sees more Z=2 particles (higher
partial cross-section) than Z=1 ones with respect to the expectations
of both FLUKA and GEANT4.

Further studies have been performed to investigate if the origin of
.such a discrepancy between Monte Carlo and data is due to some
instrumental effect of our experimental setup.  If the fragmentation
products can hits the DSSDs within few mm an overlap between charge
clusters may occurr leading to misidentified products. Since a typical
proton cluster is 2.5 strips (about 1mm) wide, if three or more
protons hits the detector within 1.5-2 mm the total product charge can
be confused with that of a helium.

To establish the frequency of occurrence of this, we performed MC
simulations and we seen that, depending on target material, a
fraction going from 1.5\% to 5\% of total Z=1 events, at least three
protons, in 1.5 mm, may overlap to mimic a helium cluster in
Thin1. The fraction variation depends on target mass and it is lowest
for graphite and highest for lead.  This effect has been included in
systematic uncertainties and, taking also into account of the
reasonable agreement in the fragment angular distribution shown in
Fig. 13, does not allow to explain the discrepancies between MC and
data we observed in Z=2 and Z=1 events.

Differences on cross-section values calculated by using different
Monte Carlo codes and physics lists are also remarkable.  Here it is
important to recall that the partial cross-section both in data
analysis and in Monte Carlo are calculated by using the highest charge
in a given event. In other words, the partial cross-sections of Z=2
are coming from events containing 3 Helium or 2 Helium and 2 protons
or 1 Helium and 4 protons. Naturally, partial cross-section of Z=1
particles are deriving from 6 proton events only.  In order to confirm
or reject the above mentioned discrepancies further studies using an
experimental setup with improved Z=1 and Z=2 separation capability may
be of help.

\clearpage

\section*{Acknowledgment}
 Authors would like to express their deepest gratitude to
 Dr. D. Schardt for his invaluable help and support for preparation
 and execution of data collection at GSI. 

We thank FLUKA and GEANT4 Collabotations for their critical reading
and comments. Authors would like to thank also ESA contract officer
Dr. A. Menicucci for her support and encouragement during the
execution of present work. Last but not least, we thank our colleagues
from Fraunhofer INT, Germany, for their cooperation in this ESA
contract.



\end{document}